\documentclass[article,aps,prd,
twocolumn,
superscriptaddress]{revtex4}
\usepackage{amssymb,amsmath,graphicx,color,bbold}
\begin{document}

\title{Linear spin-2 fields in most general backgrounds}

\author{Laura Bernard}
\affiliation{UPMC-CNRS, UMR7095, Institut d'Astrophysique de Paris, GReCO, 98bis boulevard Arago, F-75014 Paris, France}

\author{C\'edric Deffayet}
\affiliation{UPMC-CNRS, UMR7095, Institut d'Astrophysique de Paris, GReCO, 98bis boulevard Arago, F-75014 Paris, France}
\affiliation{IHES, Le Bois-Marie, 35 route de Chartres, F-91440 Bures-sur-Yvette, France}

\author{Angnis~Schmidt-May}
\affiliation{Institut f\"ur Theoretische Physik, Eidgen\"ossische Technische Hochschule Z\"urich,
Wolfgang-Pauli-Strasse 27, 8093 Z\"urich, Switzerland}

\author{Mikael von Strauss}
\affiliation{UPMC-CNRS, UMR7095, Institut d'Astrophysique de Paris, GReCO, 98bis boulevard Arago, F-75014 Paris, France}

\begin{abstract}
We derive the full perturbative equations of motion for the most general background solutions in ghost-free bimetric theory in its metric formulation. Clever field redefinitions at the level of fluctuations enable us to circumvent the problem of varying a square-root matrix appearing in the theory. This greatly simplifies the expressions for the linear variation of the bimetric interaction terms. We show that these field redefinitions exist and are uniquely invertible if and only if the variation of the square-root matrix itself has a unique solution, which is a requirement for the linearised theory to be well-defined. As an application of our results we examine the constraint structure of ghost-free bimetric theory at the level of linear equations of motion for the first time. We identify a scalar combination of equations which is responsible for the absence of the Boulware-Deser ghost mode in the theory. The bimetric scalar constraint is in general not manifestly covariant in its nature. However, in the massive gravity limit the constraint assumes a covariant form when one of the interaction parameters is set to zero. For that case our analysis provides an alternative and almost trivial proof of the absence of the Boulware-Deser ghost. Our findings generalise previous results in the metric formulation of massive gravity and also agree with studies of its vielbein version.
\end{abstract}

\maketitle

\newcommand{\be}{\begin{equation}}
\newcommand{\ee}{\end{equation}}
\def\ph{\phantom}
\newcommand{\beqn}{\begin{eqnarray}}
\newcommand{\eeqn}{\end{eqnarray}}
\newcommand{\td}{\mathrm{d}}
\newcommand{\p}{\partial}
\newcommand{\dd}{\mathrm{d}}
\newcommand{\nn}{\nonumber}
\newcommand{\Tr}{\mathrm{Tr}}
\newcommand{\gmn}{g_{\mu\nu}}
\newcommand{\fmn}{f_{\mu\nu}}
\newcommand{\bgmn}{\bar{g}_{\mu\nu}}
\newcommand{\bfmn}{\bar{f}_{\mu\nu}}

\newcommand{\ba}{\begin{eqnarray}}
\newcommand{\ea}{\end{eqnarray}}

\section{Introduction}
The study of massive spin-2 fields and nonlinear interactions between massless and massive spin-2 fields at the classical level has seen a vital resurgence in recent years. The subject has an interesting history which started already in 1939 with the work of Fierz and Pauli \cite{Fierz:1939ix}, who first obtained the consistent linear equations for both massless and massive spin-2 fields propagating on a Minkowski background. It is now known that nontrivial nonlinear interactions between massless spin-2 fields are forbidden on quite general grounds \cite{Boulanger:2000rq}. On the other hand, the main theoretical obstacle to nonlinear theories of massive spin-2 fields or, equivalently, to generalising Fierz-Pauli theory to other than Einstein backgrounds, was the generic presence of a ghost-like scalar excitation. This fatal instability was discovered in 1972 and has become known as the Boulware-Deser ghost \cite{Boulware:1973my}. Its discovery hampered the theoretical interest in massive spin-2 theories for a long period of time. 

It should be noted that, in the metric formulation, any theory of a massive spin-2 field contains two symmetric rank-2 tensors (henceforth referred to as metrics with an obvious abuse of language) since no (by definition non-derivative) Lorentz invariant mass term can be constructed out of only one metric. This means that any theory of a massive spin-2 field naturally falls into the realm of bimetric theories. In the remainder of this work, by ``massive gravity" we will mean a theory where one of these metrics is treated as non-dynamical whereas by ``bimetric" we will mean a theory where both metrics have full dynamics.

After observations confirmed the accelerated expansion of the universe \cite{Riess:1998cb,Perlmutter:1998np} and the cosmological constant problem \cite{Weinberg:1988cp} started to put increasing pressure on field theorists, much theoretical effort was spent on studying various modifications of gravity to address these issues. This included extra-dimensional scenarios such as the Dvali-Gabadadze-Porrati (DGP) model \cite{Dvali:2000hr} and generalisations thereof, geared mainly towards accomplishing self-accelerating solutions \cite{Deffayet:2000uy,Deffayet:2001pu} or a filtering of the vacuum energy (see e.g.~\cite{deRham:2007rw}). It can actually be realised that any model responsible for filtering out, or degravitating, a long wavelength mode such as a cosmological constant can effectively be described as a theory of massive gravity (with non-constant mass in general) \cite{Dvali:2007kt}. This, along with results \cite{Deffayet:2001uk} confirming that the DGP model actually realised the long conjectured Vainshtein mechanism \cite{Vainshtein:1972sx} (for a recent review on the Vainshtein mechanism see~\cite{Babichev:2013usa}) led to an increased interest in revisiting the theory of massive spin-2 fields. Based on the Yang-Mills inspired intuition proposed in \cite{ArkaniHamed:2002sp} and the subsequent analysis of \cite{Creminelli:2005qk, Deffayet:2005ys}, in 2010 a major breakthrough in the developments was achieved when de Rham, Gabadadze and Tolley (henceforth dRGT) decided to repeat the analysis of \cite{Creminelli:2005qk}, corrected a simple sign mistake of that work\footnote{This unfortunate sign mistake was actually due to copying a very basic equation of \cite{ArkaniHamed:2002sp}, where the sign was not important for the arguments made there.} and were able to find a nonlinear action which was devoid of the Boulware-Deser ghost in a decoupling limit regime of the parameters, where only the longitudinal mode of the massive spin-2 field is considered~\cite{deRham:2010ik,deRham:2010kj}. 
As expected, the dRGT theory correctly linearised to the Fierz-Pauli form for a massive spin-2 fluctuation on Minkowski backgrounds. 

The decoupling limit analysis was an important first step towards finding a consistent nonlinear theory of a massive spin-2 field since it provided a consistency check for a necessary but not sufficient criterion for any consistent theory to fulfil.

The short-comings of the decoupling limit analysis of dRGT were however soon criticized (see e.g.~\cite{Alberte:2010qb,Chamseddine:2011mu} but also~\cite{deRham:2011rn,Hassan:2012qv} for counter arguments). The next major breakthrough developments came when Hassan and Rosen managed to demonstrate via a nonlinear ADM analysis that the theory initially suggested by dRGT only propagated 5 degrees of freedom when the non-dynamical metric was taken to be flat~\cite{Hassan:2011vm,Hassan:2011hr} (in line with the structure suggested by~\cite{deRham:2010ik,deRham:2010kj})


Due to the powerful nature of the reformulation by Hassan and Rosen they shortly managed to extend their proof to the case when the non-dynamical metric was taken to be completely arbitrary \cite{Hassan:2011tf} (see also~\cite{Hassan:2012qv}). Subsequently they also showed that the proof could be extended to the bimetric case where both metrics are dynamical and that the theory then propagated the $2+5=7$ degrees of freedom of a massless spin-2 field interacting nonlinearly with a massive spin-2 field \cite{Hassan:2011zd,Hassan:2011ea}. From a field theoretical perspective this increase in degrees of freedoms makes perfect sense since the bimetric extension is fully covariant under the diagonal group of common diffeomorphisms of the two metrics and there should be a massless spin-2 field associated to this gauge invariance. This theory provides a very interesting, and from a theoretical perspective minimal, extension of Einstein's theory of general relativity. The dynamical nature of the theory is quite restrictive and also seems to imply that it avoids many of the potential drawbacks present in its massive gravity version, see e.g.~\cite{Fasiello:2013woa,DeFelice:2014nja,Hassan:2014vja,Akrami:2015qga}.

In the present work we start out from the ghost-free bimetric theory obtained by Hassan and Rosen (see~\cite{Schmidt-May:2015vnx}~for a recent review). In the metric formulation the theory contains a square-root matrix and is difficult to handle in general and in particular when considering perturbations. While there exist a vielbein formulation of the theory \cite{Hinterbichler:2012cn} which avoids the presence of a square-root matrix the two theories are in general not equivalent without further restrictions \cite{Deffayet:2012zc,Banados:2013fda,Hassan:2014gta}. In fact, exactly when the vielbein theory does not have a metric description it seems to be again plagued by a ghost \cite{deRham:2015cha}. In this work we shall only work within the metric formulation of the theory. In the recent works \cite{Bernard:2014bfa,Bernard:2015mkk} three of us managed to resolve the problem of computing the variation of the square-root matrix explicitly for the first time within the massive gravity context but for general backgrounds. In these works it was found that the backgrounds had to meet a certain criterion in order for the variation of the square-root to have a unique solution and hence allow for a well-defined linearised problem (to wit:~the spectrum of eigenvalues of the square-root matrix and its negative must not intersect). This was then used to obtain, for a subset of dRGT massive gravities linearised on an arbitrary background, a fully covariant constraint responsible for the removal of the Bouware-Deser ghost. We hence provided the first complete and covariant form of a theory for a massive graviton propagating on a totally arbitrary background metric with 5 (or less) polarisations. The obtained theory involves a complicated mass term which depends on the curvature of the background in a highly non trivial way. Here we first extend these results to provide the explicit linearised equations of motion in the fully dynamical bimetric theory for general backgrounds. In doing so, we consider field redefinitions at the level of perturbations which allow us to circumvent the difficulty of linearising the square-root matrix. These redefinitions are shown to be uniquely invertible exactly when the variation of the square-root matrix exists and the linearised problem is well-defined to start with.

The field redefinitions and the expressions we provide can be of use for any perturbative analysis within the bimetric framework. 
In this work, as an immediate application of having obtained the linearised field equations in a simplified form, we study for the first time the structure of the Lagrangian constraints of the bimetric equations, extending our work \cite{Bernard:2014bfa,Bernard:2015mkk} on massive gravity. We are able to find a scalar constraint responsible for the removal of the Boulware-Deser ghost mode. For general parameters, the constraint we find is not of a manifestly covariant form, in the sense that it is not devoid of second-order covariant derivatives. Rather, one has to closely examine the component form in order to see that the expression we derive in fact does constitute a constraint on the dynamical variables. In the massive gravity limit of the theory, where one of the metrics is taken to be non-dynamical, a similar conclusion holds for the general case. However, if a certain interaction parameter is set to zero then for arbitrary values of the remaining parameters the constraint becomes manifestly covariant. This conclusion agrees with previous studies of the massive gravity version of the theory in the vielbein formulation \cite{Deffayet:2012nr,Deser:2014hga} and in the metric formulation \cite{Bernard:2014bfa,Bernard:2015mkk}, and with our analysis we shed more light on these results.

\vspace{0.5cm}

The remainder of this article is organised as follows. In section \ref{sec: review} we review the basic structure of ghost-free bimetric theory and its cousin massive gravity. We also discuss some general aspects of linearisation of the theory and the problems that arise. In section \ref{sec: redef} we motivate and introduce a redefined set of fluctuation variables which are tailored towards overcoming the main obstacles of linearisation. Utilising these variables we then obtain the simplified linearised equations of motion in section \ref{sec: linearisation}. As an application we first restrict to massive gravity and perform a covariant constraint analysis in section \ref{sec: covconstr}, which both confirms and generalises previous results. Afterwards we return to the general bimetric case and discuss the constraint analysis in that setup. We end by discussing our results in section \ref{sec: conclusions}.

\vspace{0.5cm}

\paragraph{A note on notation:} We warn the reader that the presence of two metrics and expressions with many indices will sometimes force us to use different metrics to raise and contract indices. In order to avoid confusion, we will always move indices of quantities appearing in the field equations of the metric $g_{\mu \nu}$ (respectively of the metric $f_{\mu \nu}$)
with the metric $g_{\mu \nu}$ (respectively with the metric $f_{\mu \nu}$). Moreover, we will use a $\sim$ on top of quantities that are defined with respect to $\fmn$ to distinguish them from quantities defined with respect to $\gmn$. For example $V^{\mu\nu}$ will mean $g^{\mu\rho}g^{\nu\sigma}V_{\rho\sigma}$, while $\tilde{V}^{\mu\nu}$ will mean $f^{\mu\rho}f^{\nu\sigma}\tilde{V}_{\rho\sigma}$ etc. Whenever any confusion may arise we will avoid such index raisings and contractions and keep our expressions explicit. Curvatures are defined according to the rule $\left[\nabla_\mu,\nabla_\nu\right]\omega_\rho=R_{\mu\nu\rho}^{\ph{\mu\nu\rho}\sigma}\omega_\sigma$ and $R_{\mu\nu}=R_{\mu\sigma\nu}^{\ph{\mu\sigma\nu}\sigma}$ with respect to the covariant derivatives of either metric.

\section{Review of bimetric theory \& massive gravity}\label{sec: review}
The ghost-free bimetric theory is defined by the action \cite{Hassan:2011zd}
\begin{align}\label{bgaction}
S=m_g^2\int\td^4x\biggl[&\sqrt{|g|}R(g)+\alpha^2\sqrt{|f|}R(f)\nn\\
&\quad\quad-2m^2\sqrt{|g|}V\left(S;\beta_n\right)\biggr]\,,
\end{align}
where $m_g$ is the generalised "Planck mass scale", $\alpha^2$ measures the relative kinetic strengths for the two metrics and $m$ sets the scale of the spin-2 mass. In addition, the theory contains 5 dimensionless interaction parameters $\beta_n$, where $\beta_0$ and $\beta_4$ act as bare cosmological constants for $\gmn$ and $\fmn$, respectively, and hence encode nonlinear self-interactions while $\beta_1,\beta_2,\beta_3$ measure proper nonlinear interactions between the metrics. The precise form of the interaction potential $V$ is determined by demanding absence of the Boulware-Deser ghost and is given by~\cite{deRham:2010kj, Hassan:2011zd}
\be\label{Vdef}
V\left(S;\beta_n\right) = \sum_{n=0}^4\beta_n e_n(S)\,,
\ee
where the $e_n(S)$ are the elementary symmetric polynomials defined in terms of the eigenvalues of the matrix $S$. They can be constructed iteratively through the recursive relation (starting from $e_0(S)=1$)
\be\label{endef}
e_n(S) = -\frac{1}{n}\sum_{k=1}^{n}(-1)^{k}\Tr[S^k]e_{n-k}(S)\,,\qquad n\geq1\,.
\ee
Here $\Tr[S^k]=S^{\mu_1}_{~\mu_2}S^{\mu_2}_{~\mu_3}\cdots S^{\mu_k}_{~\mu_1}$ is understood as the matrix trace of the $k$th power of the tensor $S$ considered as a matrix. Note that $e_4(S)=\det(S)$ and $e_n(S)=0$ for all $n>4$. Finally, the matrix argument $S$ is a square-root matrix defined through the relation $S^\rho_{~\sigma}S^\sigma_{~\nu}=g^{\rho\mu}\fmn$. 

Whenever the inverse $S^{-1}$ exists,\footnote{We will assume invertibility of $S$ throughout since this follows from the invertibility of both $\gmn$ and $\fmn$, which must be assumed for the bimetric theory to be well defined.} the identity
\be\label{enSSI}
e_n(S^{-1})=\frac{e_{4-n}(S)}{e_4(S)}\,,
\ee
can be used to see that the theory treats $\gmn$ and $\fmn$ in a completely symmetric fashion. In fact, in the absence of matter couplings the theory is invariant under the discrete interchanges
\be\label{intsym}
\alpha^{-1}\gmn\leftrightarrow\alpha\fmn\,,\qquad
\alpha^{4-n}\beta_n\leftrightarrow \alpha^n\beta_{4-n}\,.
\ee
This has often been used to set $\alpha=1$ by scaling e.g.~$\fmn$ together with the $\beta_n$, which can be done consistently if $\fmn$ does not couple to any source. As pointed out in \cite{Akrami:2015qga}, care should be taken when one considers a perturbative expansion after such a rescaling.

The variation of the $e_n(S)$, under $\delta S$ of $S$, as defined in \eqref{endef} is given by\footnote{For an explicit derivation of this see e.g.~\cite{Bernard:2015mkk}.}
\be\label{d_e_n}
\delta e_n(S)=-\sum_{k=1}^n(-1)^k\Tr[S^{k-1}\delta S]\,e_{n-k}(S)\,,\quad n\geq1\,,
\ee
with $\delta e_0(S)=0$ (since $e_0(S)=1$). Using this together with $2\Tr[S^{k-1}\delta S]=\Tr[S^{k-2}\delta S^2]$ and\footnote{We will frequently use the notation $\delta S^2=\delta(S^2)$. Since we never go beyond linear perturbation theory no confusion should arise from this.} $\delta S^2=-g^{-1}\delta g S^2 + g^{-1}\delta f$ (which follows from $S^2=g^{-1}f$) it is straightforward to derive the vacuum field equations of the theory
\begin{subequations}\label{eombim}
\begin{align}
E_{\mu\nu}&\equiv\mathcal{G}_{\mu\nu}+m^2V_{\mu\nu}=0\,,\quad
V_{\mu\nu}\equiv -\frac{2}{\sqrt{|g|}}\frac{\p(\sqrt{|g|}V)}{\p g^{\mu\nu}} \label{bgeoms_g}\\
\tilde{E}_{\mu\nu}&\equiv \tilde{\mathcal{G}}_{\mu\nu}+\frac{m^2}{\alpha^2}\tilde{V}_{\mu\nu}=0\,,\quad
\tilde{V}_{\mu\nu}\equiv -\frac{2}{\sqrt{|f|}}\frac{\p(\sqrt{|g|}V)}{\p f^{\mu\nu}}\,.
\label{bgeoms_f}
\end{align}
\end{subequations}
Here $\mathcal{G}_{\mu\nu}=R_{\mu\nu}-\tfrac1{2}\gmn R$ is the Einstein tensor computed with respect to $\gmn$ and $\tilde{\mathcal{G}}_{\mu\nu}=\tilde{R}_{\mu\nu}-\tfrac1{2}\fmn \tilde{R}$ is the Einstein tensor computed with respect to $\fmn$. Note that the second line can be obtained directly from the first by making use of the interchange symmetry \eqref{intsym}. The interaction contributions are matrix polynomials in $S$ and are given explicitly by
\begin{align}\label{Vmndef}
V_{\mu\nu}&=g_{\mu\rho}\sum_{n=0}^3(-1)^n\beta_nY^\rho_{(n)\,\nu}(S)\,,\\
\tilde{V}_{\mu\nu}&=f_{\mu\rho}\sum_{n=0}^3(-1)^n\beta_{4-n}Y^\rho_{(n)\,\nu}(S^{-1})\,,
\end{align}
where the tensors $Y_{(n)}(S)$ are defined as
\be
Y^\rho_{(n)\,\nu}(S)=\sum_{k=0}^n(-1)^ke_k(S)[S^{n-k}]^\rho_{~\nu}\,.
\ee
For example, written out in full we have that
\begin{align}
V_{\mu\nu}=g_{\mu\rho}\biggl[&\beta_0\delta^\rho_{\nu}-
\beta_1\left(S^\rho_{~\nu}-e_1\delta^\rho_\nu\right)\nn\\
&+\beta_2\left([S^2]^\rho_{~\nu}-e_1S^\rho_{~\nu}+e_2\delta^\rho_\nu\right)\nn\\
&-\beta_3\left([S^3]^\rho_{~\nu}-e_1[S^2]^\rho_{~\nu}+e_2S^\rho_{~\nu}-e_3\delta^\rho_\nu\right)\biggr]\,.
\end{align}
We note that $V_{\mu\nu}$ and $\tilde{V}_{\mu\nu}$ as written in \eqref{Vmndef} are symmetric in their indices, albeit not manifestly. Let us define $gS$ as the covariant tensor $\left(gS\right)_{\mu \nu} = g_{\mu \rho} S^{\rho}_{~\nu}$, similarly $fS^{-1}$ as 
$\left(fS^{-1}\right)_{\mu \nu} = f_{\mu \rho} (S^{-1})^{\rho}_{~\nu}$
 and denote the matrix transpose operation by~$^T$. Then indeed we have that $(gS)^{\mathrm{T}}=gS=fS^{-1}=(fS^{-1})^{\mathrm{T}}$, along with similar identities showing that $S$ and $S^{-1}$ are in fact symmetric whenever their indices are raised or lowered using either of $\gmn$ or $\fmn$.\footnote{These identities can easily be proven either by a formal expansion of the square-root as in \cite{Hassan:2012wr} or by explicit matrix manipulations as in \cite{Baccetti:2012re}.}

General covariance of the theory under the diagonal group of common diffeomorphisms implies the divergence identities (see e.g.~\cite{Damour:2002ws})
\be\label{covID_1}
\sqrt{|g|}\,g^{\mu\rho}\nabla_\rho V_{\mu\nu} = - \sqrt{|f|}\,f^{\mu\rho}\tilde{\nabla}_\rho \tilde{V}_{\mu\nu}\,,
\ee
as well as the algebraic identities \cite{Hassan:2014vja} (see also~\cite{Baccetti:2012bk, Volkov:2012zb})
\be\label{covID_2}
\sqrt{|g|}\,g^{\rho\mu}V_{\mu\nu}+\sqrt{|f|}\,f^{\rho\mu}\tilde{V}_{\mu\nu}-\sqrt{|g|}\,V\delta^\rho_\nu
=0\,,
\ee
where $V$ is the interaction potential \eqref{Vdef} appearing in the action \eqref{bgaction}. We stress that these identities are a direct consequence of covariance and are of little use if one does not treat both metrics dynamically (clearly $\tilde{V}_{\mu\nu}$ is of no relevance if $\fmn$ is not treated dynamically). The identities \eqref{covID_2} can be used together with the bimetric equations \eqref{eombim} to prove that if either one of the metrics describes an Einstein space then the other one must also be an Einstein space \cite{Hassan:2014vja} (see also~\cite{Blas:2005yk}), a conditional identity which is lost in the massive gravity version of the theory where one usually considers the fixed metric to be either flat or of constant curvature while the other one is not.

\subsection{The massive gravity limit}\label{sec: MGlim}
For later reference we discuss briefly the massive gravity limit of bimetric theory at the level of the equations of motion \cite{Baccetti:2012bk,Hassan:2014vja}. Since the theory treats the two metrics symmetrically we must of course make a choice in what we mean by a massive gravity limit. Here we will mean the limit in which the equations for $\gmn$ becomes the equations for a nonlinear massive spin-2 field. Furthermore, since we will generically be studying the vacuum equations we consider a restricted limit here with no matter sources for either metric.\footnote{This will lead to $\fmn$ being an Einstein spacetime in the limit. More generally one can consider including some exotic matter source for $\fmn$ with an appropriate $\alpha$ dependence in the matter couplings in order to generate more general $\fmn$ metrics in the limit.} The massive gravity limit can then be achieved by considering the limit $\alpha\rightarrow\infty$. From the bimetric equations of motion
\be\label{bgeoms_gf}
\mathcal{G}_{\mu\nu}+m^2V_{\mu\nu}=0\,,\qquad
\tilde{\mathcal{G}}_{\mu\nu}+\frac{m^2}{\alpha^2}\tilde{V}_{\mu\nu}=0\,,
\ee
it is clear that, if we want to keep nontrivial interactions in the $\gmn$ equations after this limit is taken, the only $\beta_i$ parameter that can be rescaled in an $\alpha$ dependent way while taking the limit  $\alpha\rightarrow\infty$ is $\beta_4$.
This is because, while $\beta_1,\beta_2,\beta_3$ and $\beta_4$ all appear in $\tilde{V}_{\mu\nu}$, only $\beta_4$ does not appear in $V_{\mu\nu}$. Hence any scalings of $\beta_1,\beta_2$ or $\beta_3$ are not allowed since this would cause $V_{\mu\nu}$ to diverge.
Assuming therefore that none of the other $\beta_n$ are affected by this limit and taking $\beta_4=\alpha^2\Lambda_f/m^2$, then $\alpha\rightarrow\infty$ implies
\be\label{bgeoms_MGlim}
\mathcal{G}_{\mu\nu}+m^2V_{\mu\nu}=0\,,\qquad
\tilde{\mathcal{G}}_{\mu\nu}+\Lambda_f \fmn=0\,.
\ee
Hence $\fmn$ is a constant curvature solution while $\gmn$ obeys the equations obtained from a nonlinear massive gravity action\footnote{There is no covariant action which can reproduce both of the equations in \eqref{bgeoms_MGlim}. It is however possible to take the limit $\alpha\rightarrow\infty$ in the action such that the quadratic fluctuations of a background solution $\fmn$ decouple (see e.g.~\cite{deRham:2014zqa}). This results in a non-covariant action and works provided that the background solution is fully well-behaved in the limit, which is however not always the case \cite{Hassan:2014vja, Zhang:2014wia}. If treated correctly these limiting procedures are of course consistent with each other.} (incidentally the bimetric action \eqref{bgaction} with $\alpha=0$ and $\fmn$ treated as non-dynamical). As discussed in \cite{Hassan:2014vja} the massive gravity limit of the bimetric theory is quite subtle and requires some care, but at least a subset of bimetric solutions do admit such a limit, whether the limit is taken in the equations of motion as here or directly in the action. At the level of linear fluctuations around such well-behaved backgrounds one can then consider canonically normalised fluctuations $\delta\gmn/m_g$ and $\delta\fmn/(\alpha m_g)$ such that also the linear theory coincides with the standard treatment of massive gravity, where $\fmn$ is kept fixed and its fluctuations are ignored. Later on when we discuss covariant constraints of the massive gravity equations we do so at the level of linear perturbations and therefore we will ignore potential nonlinear problems that may arise in taking the limit and simply treat the equations as well-defined at the level we work with them.

As remarked upon in the introduction of this subsection, completely analogous statements can of course be made in the limit $\alpha\rightarrow0$ where instead $\fmn$ will obey the nonlinear equations of massive gravity and $\gmn$ is a constant curvature metric with $\beta_0$ scaled appropriately.

\subsection{Linearised theory of fluctuations}\label{sec: linpert}
We will now turn to the general problem of linearising the field equations \eqref{eombim}. To this end we expand the two metrics around some background solutions of the equations of motion as follows
\be
\gmn\rightarrow\gmn+\delta\gmn\,,\qquad\fmn\rightarrow\fmn+\delta\fmn\,,
\ee
where, as our notation suggests, we retain the label $\gmn$ and $\fmn$ for the background solutions to avoid unnecessary clutter. The equations of motion for the perturbations then read, schematically,
\begin{subequations}\label{lineoms}
\begin{align}
\delta E_{\mu\nu}&=\,\delta\mathcal{G}_{\mu\nu}+m^2\delta V_{\mu\nu}=0\,,\label{bgeoms_dg} \\
\delta\tilde{E}_{\mu\nu}&=\, \delta\tilde{\mathcal{G}}_{\mu\nu}
+\frac{m^2}{\alpha^2}\delta\tilde{V}_{\mu\nu}=0\,.
\label{bgeoms_df}
\end{align}
\end{subequations}
The kinetic terms of the equations of motion are straightforward to obtain as they are simply the linearised Einstein tensors. In the $\gmn$ equations they read
\begin{align}
\label{dGmn}
\delta\mathcal{G}_{\mu\nu}
&=\mathcal{E}_{\mu\nu}^{\ph\mu\ph\nu\rho\sigma}\delta g_{\rho\sigma}+\tfrac1{2}\left[
g_{\mu\nu} R^{\rho\sigma}-\delta^\rho_\mu \delta^\sigma_\nu R\right]\delta g_{\rho\sigma}\,,
\end{align}
where, for later considerations, we have defined the explicit 2-derivative operator
\begin{align}\label{gEinstOp}
\mathcal{E}_{\mu\nu}^{\ph\mu\ph\nu\rho\sigma}\delta g_{\rho\sigma}
\equiv &-\tfrac1{2}\biggl[\delta^\rho_\mu\delta^\sigma_\nu\nabla^2+g^{\rho\sigma}\nabla_\mu\nabla_\nu 
-\delta^\rho_\mu\nabla^\sigma\nabla_\nu\nn\\
&-\delta^\rho_\nu\nabla^\sigma\nabla_\mu
-g_{\mu\nu} g^{\rho\sigma}\nabla^2 +g_{\mu\nu}\nabla^\rho\nabla^\sigma\biggr]\delta g_{\rho\sigma}\,.
\end{align}
Observe that in the above expressions we have used $\gmn$ to raise and contract indices. As mentioned already in the introduction, in order to avoid unnecessarily lengthy expressions we will sometimes adopt this convention whenever an expression only contains quantities defined with respect to $\gmn$ and its equations as above. Analogously we will use $\fmn$ to raise and contract indices in expressions only containing quantities defined with respect to $\fmn$ and its equations, as in the following expressions. Whenever any confusion may arise we will keep all indices and their placings explicit. In the $\fmn$ equations the kinetic contributions similarly read
\begin{align}
\label{dtGmn}
\delta\tilde{\mathcal{G}}_{\mu\nu}
&=\tilde{\mathcal{E}}_{\mu\nu}^{\ph\mu\ph\nu\rho\sigma}\delta f_{\rho\sigma}+\tfrac1{2}\left[
f_{\mu\nu} \tilde{R}^{\rho\sigma}-\delta^\rho_\mu \delta^\sigma_\nu \tilde{R}\right]\delta f_{\rho\sigma}\,,
\end{align}
where analogously
\begin{align}\label{fEinstOp}
\tilde{\mathcal{E}}_{\mu\nu}^{\ph\mu\ph\nu\rho\sigma}\delta f_{\rho\sigma}
\equiv &-\tfrac1{2}\biggl[\delta^\rho_\mu\delta^\sigma_\nu\tilde{\nabla}^2
+f^{\rho\sigma}\tilde{\nabla}_\mu\tilde{\nabla}_\nu 
-\delta^\rho_\mu\tilde{\nabla}^\sigma\tilde{\nabla}_\nu\nn\\
&-\delta^\rho_\nu\tilde{\nabla}^\sigma\tilde{\nabla}_\mu
-f_{\mu\nu} f^{\rho\sigma}\tilde{\nabla}^2 +f_{\mu\nu}\tilde{\nabla}^\rho\tilde{\nabla}^\sigma\biggr]\delta f_{\rho\sigma}\,,
\end{align}
The linearisation of the interaction contributions \eqref{Vmndef} requires some further work but using \eqref{d_e_n} together with (using matrix notation within the square brackets)
\be
\delta \left[S^{n-k}\right]^\rho_{~\nu}
=\sum_{m=0}^{n-k-1} \left[S^{m}\delta S\, S^{n-k-m-1}\right]^\rho_{~\nu}\,,
\ee
which is a direct consequence of applying the chain rule, it is a simple matter of algebra to arrive at the following intermediate results
\begin{widetext}
\begin{align}\label{dVmn}
\delta V_{\mu\nu} = g^{\rho\sigma}V_{\sigma\nu}\delta g_{\mu\rho}
&-g_{\mu\rho}\sum_{n=1}^{3}(-1)^{n}\beta_n
\sum_{k=1}^{n}(-1)^{k}\biggl\{\left[S^{n-k}\right]^{\rho}_{~\nu}
\sum_{m=1}^{k}(-1)^{m}e_{k-m}(S)\left[S^{m-1}\delta S\right]^{\sigma}_{~\sigma}\nn\\
&\qquad\qquad\qquad\qquad\qquad\qquad
+e_{k-1}(S)\sum_{m=0}^{n-k}\left[S^{m}\delta S S^{n-k-m}\right]^{\rho}_{~\nu}
\biggr\}\,,
\end{align}
and
\begin{align}\label{dtVmn}
\delta \tilde{V}_{\mu\nu} = f^{\rho\sigma}\tilde{V}_{\sigma\nu}\delta f_{\mu\rho}
&+f_{\mu\rho}\sum_{n=1}^{3}(-1)^{n}\beta_{4-n}
\sum_{k=1}^{n}(-1)^{k}\biggl\{\left[S^{k-n}\right]^{\rho}_{~\nu}
\sum_{m=1}^{k}(-1)^{m}e_{k-m}(S^{-1})\left[S^{-m-1}\delta S\right]^{\sigma}_{~\sigma}\nn\\
&\qquad\qquad\qquad\qquad\qquad\qquad
+e_{k-1}(S^{-1})\sum_{m=0}^{n-k}\left[S^{-m-1}\delta S S^{m+k-n-1}\right]^{\rho}_{~\nu}
\biggr\}\,.
\end{align}
\end{widetext}
These expressions should be symmetrised over the $\mu\nu$ indices, but when evaluated on any background solutions, the full equations of motion will actually be symmetric even with the given form when the above expressions are combined with the kinetic terms. We note that the first terms of these expressions, i.e.~$g^{\rho\sigma}V_{\sigma\nu}\delta g_{\mu\rho}$ and $f^{\rho\sigma}\tilde{V}_{\sigma\nu}\delta f_{\mu\rho}$, come from the variation of the volume element. For the proportional background solutions, $\fmn=c^2\gmn$, they encode purely cosmological contributions (i.e.~non-mass contributions). Using the background relations $\mathcal{G}_{\sigma\nu}+m^2V_{\sigma\nu}=0$ and $\alpha^2\tilde{\mathcal{G}}_{\sigma\nu}+m^2\tilde{V}_{\sigma\nu}=0$ these can be added up to the corresponding terms from the kinetic operators in \eqref{dGmn} and \eqref{dtGmn}, respectively. The terms contained within the sums on the other hand constitute the non-minimal interactions which render a combination of the fluctuations massive.

In these sums the last terms, on the second lines, present a technical problem\footnote{The first terms containing the variation $\delta S$ under a trace present no problem since $2\Tr[S^{m-1}\delta S]=\Tr[S^{m-2}\delta S^2]$ and $\delta S^2$ is easy to obtain.} since they contain the variation of a square-root matrix, i.e.~$\delta S = \delta\sqrt{g^{-1}f}$. This variation had been computed in some simple examples, e.g.~for black holes and in a cosmological context~\cite{Comelli:2012db, Berg:2012kn, Kodama:2013rea, Kobayashi:2015yda}, but it was only recently in \cite{Bernard:2014bfa,Bernard:2015mkk} that the general variation was computed for the first time (see also~\cite{Guarato:2013gba} which was however not fully general). These works utilised that the problem could be stated, in matrix notation,
\be\label{dS2sylvester}
S\delta S + \delta S S = \delta S^2\,,
\ee
where the variation $\delta S^2 = -g^{-1}\delta g S^{2}+g^{-1}\delta f$ is known. The realisation that this is a type of Sylvester matrix equation allows to use results known from the mathematical literature \cite{Sylvesterpoly} to obtain the solution as\footnote{Note that in \cite{Bernard:2015mkk} an alternative derivation and form for the solution of $\delta S$ was also provided and shown to be equivalent to the one presented here.}
\be \label{sylvestersol}
\delta S=\frac1{2}\mathbb{X}^{-1}\sum_{k=1}^4\sum_{m=0}^{k-1}
(-1)^me_{4-k}(S)S^{k-m-2}\delta S^2S^m\,,
\ee
where $\mathbb{X}=e_3(S)\mathbb{1}+e_1(S)S^2$. This solution, which we stress is exact and linear in $\delta S^2$, exists and is unique if and only if the spectrum $\sigma(S)$ of $S$ (i.e.~the set of eigenvalues of $S$) and the spectrum of $-S$ do not intersect, i.e.~$\sigma(S)\cap\sigma(-S)=\emptyset$, which is in fact equivalent to the statement that $\mathbb{X}$ is invertible \cite{Bernard:2015mkk}. This implies that unless one imposes this external condition the theory cannot be unambiguously linearised. This condition is generically satisfied for physically interesting solutions of the theory (for example if $S$ has distinct or only positive eigenvalues) so we will assume that the background solutions are such that this condition is not violated.\footnote{This condition is correlated to the requirement of having a real square-root (see e.g.~\cite{Deffayet:2012zc}) but is slightly stronger and in fact connects to requirements of also having a locally well posed initial value problem \cite{fawad_mikica}.} When $\mathbb{X}$ is invertible one can use the Cayley-Hamilton theorem (cf.~\eqref{CHthrm}) to show that it can be obtained by the expression
\be\label{Xinv}
\mathbb{X}^{-1}=\frac{\left(e_3-e_1e_2\right)\mathbb{1}+e_1^2S-e_1S^2}{e_1^2e_4+e_3^2-e_1e_2e_3}\,,
\ee
where all the $e_n$ are now functions of $S$ (see~\cite{Bernard:2015mkk} for details). 

In principle it is now possible to combine all of the results of this section and obtain the equations of motion for the perturbations on any given set of background solutions. The resulting expressions are however quite lengthy and not always easy to analyse in practise, mostly because the solution \eqref{sylvestersol} typically contain 10 different terms multiplying $\mathbb{X}^{-1}$. Later on, in section \ref{sec: redef}, we will therefore make use of field redefinitions to simplify the problem and show that these field redefinitions exist and are invertible precisely under the same conditions for which the variation $\delta S$ exists and is unique. Before doing so we will continue to review some further aspects of the theory which will be useful for later purposes.

\subsubsection{Einstein solutions \& Mass eigenstates}\label{sec:Einstein}
With the linearised equations of motion at hand we can verify our results so far for a particularly simple set of background solutions, namely the proportional solutions for which $\fmn=c^2\gmn$. Inserting this ansatz into the background equations \eqref{eombim}, these equations reduce to two copies of the vacuum Einstein equations
\be
\mathcal{G}_{\mu\nu}+\Lambda_g\gmn=0\,,\qquad
\mathcal{G}_{\mu\nu}+\Lambda_f\gmn=0\,,
\ee
where $\Lambda_g=m^2\left(\beta_0+3c\beta_1+3c^2\beta_2+c^3\beta_3\right)$ and $\Lambda_f=\tfrac{m^2}{\alpha^2c^2}\left(c\beta_1+3c^2\beta_2+3c^3\beta_3+c^4\beta_4\right)$. Consistency of these equations of course requires that $\Lambda_g=\Lambda_f$, a condition which can be written as a quartic polynomial equation for $c$ with coefficients dependent on $\alpha$ and $\beta_n$ and hence determines $c=c(\alpha,\beta_n)$. This condition then completely specifies the backgrounds in terms of the parameters of the theory. For these backgrounds we also have the simple relations
\begin{align}\label{prop_dS}
S^{\rho}_{~\nu}&=c\,\delta^{\rho}_{\nu}\,,\qquad
e_n(S)=c^n{4\choose n}\,,\nn\\
2c\,\delta S^{\rho}_{~\nu}&=g^{\rho\mu}\left(\delta \fmn-c^2\delta \gmn\right)\,,
\end{align}
where ${4\choose n}=4!/n!(4-n)!$ is the standard binomial coefficient. The kinetic terms \eqref{dGmn} and \eqref{dtGmn} then become
\be
\delta\mathcal{G}_{\mu\nu}=
\mathcal{E}_{\mu\nu}^{\ph\mu\ph\nu\rho\sigma}\delta g_{\rho\sigma}
+\tfrac{\Lambda_g}{2}\left[
g_{\mu\nu} g^{\rho\sigma}\delta g_{\rho\sigma}-4\delta\gmn\right]\,,
\ee
and
\be
\delta\tilde{\mathcal{G}}_{\mu\nu}=
\tfrac1{c^{2}}{\mathcal{E}}_{\mu\nu}^{\ph\mu\ph\nu\rho\sigma}\delta f_{\rho\sigma}
+\tfrac{\Lambda_f}{2c^2}\left[
g_{\mu\nu} g^{\rho\sigma}\delta f_{\rho\sigma}-4\delta\fmn\right]\,,
\ee
while the linearised interaction terms \eqref{dVmn} and \eqref{dtVmn} reduce to
\begin{align}
\delta V_{\mu\nu}=&\frac{\Lambda_g}{m^2}\delta\gmn\nn\\
&+\left(\beta_1+2c\beta_2+c^2\beta_3\right)\left(\gmn\delta S^\rho_{~\rho}
-g_{\mu\rho}\delta S^\rho_{~\nu}\right)\,,
\end{align}
and
\begin{align}
\delta \tilde{V}_{\mu\nu}=&\frac{\alpha^2\Lambda_f}{c^2m^2}\delta\fmn\nn\\
&-\frac{1}{c^2}\left(\beta_1+2c\beta_2+c^2\beta_3\right)\left(\gmn\delta S^\rho_{~\rho}
-g_{\mu\rho}\delta S^\rho_{~\nu}
\right)\,.
\end{align}
Using the consistency condition $\Lambda_g=\Lambda_f\equiv\Lambda$ we then find, by considering the linear combination $\delta E_{\mu\nu}+\alpha^2c^2\delta\tilde{E}_{\mu\nu}$ of the equations of motion \eqref{lineoms}, that the combined field $\delta G_{\mu\nu}=\delta\gmn+\alpha^2\delta\fmn$ satisfies the linearised Fierz-Pauli equations of a massless spin-2 field
\begin{align}\label{FPeqsMeq0}
{\mathcal{E}}_{\mu\nu}^{\ph\mu\ph\nu\rho\sigma}\delta G_{\rho\sigma}
-\Lambda\left(\delta G_{\mu\nu}-\frac1{2}\gmn g^{\rho\sigma}\delta G_{\rho\sigma}\right)=0\,.
\end{align}
Similarly, by considering the linear combination $(c/2)\delta\tilde{E}_{\mu\nu}-(c/2)\delta E_{\mu\nu}$ we find that $\delta M_{\mu\nu}=g_{\mu\rho}\delta S^\rho_{~\nu}$ satisfies the Fierz-Pauli equations of a massive spin-2 field
\begin{align}\label{FPeqsMneq0}
{\mathcal{E}}_{\mu\nu}^{\ph\mu\ph\nu\rho\sigma}\delta M_{\rho\sigma}&
-\Lambda\left(\delta M_{\mu\nu}-\frac1{2}\gmn g^{\rho\sigma}\delta M_{\rho\sigma}\right)\nn\\
&+\frac{m_{\mathrm{FP}}^2}{2}\left(\delta M_{\mu\nu}
-\gmn g^{\rho\sigma}\delta M_{\rho\sigma}\right)=0\,,
\end{align}
where $m_{\mathrm{FP}}^2=m^2\left(\frac{1+\alpha^2c^2}{\alpha^2c^2}\right)\left(c\beta_1+2c^2\beta_2+c^3\beta_3\right)$. This of course agrees with the analysis of \cite{Hassan:2012wr} (see also~\cite{Deffayet:2011uk, Hassan:2012rq}) and mostly serves as a simple consistency check of our results up to this point but also illustrates the mass spectrum of the theory around constant curvature backgrounds.

\section{Redefined fluctuation variables}\label{sec: redef}
As outlined in section \ref{sec: linpert}, the computation of the linearised equations of motion around general background solutions is complicated due to the presence of the square-root matrix $S=\sqrt{g^{-1}f}$. The variation $\delta S$ can be obtained explicitly using either the Sylvester form given in \eqref{sylvestersol} or by using the Cayley-Hamilton theorem \cite{Bernard:2015mkk}, but the resulting expressions are lengthy and difficult to analyse analytically. Here we will employ a clever redefinition of the dynamical variables in a spirit similar to that of \cite{Schmidt-May:2014xla}. This allows us to avoid evaluating the variation of the square-root matrix explicitly and to study the linearised equations in a more compact form.

In order to replace say $\delta\gmn$ by a new fluctuation variables $\delta\gmn'$ which simplifies the expressions for the linearised equations, we define (the important issue of the invertibility of this definition is discussed below)
\begin{align}\label{dgp_def1}
\delta\gmn &\equiv\left.\frac{\delta \gmn}{\delta (S^{-1})^\rho_{~\sigma}}\right|_f f^{\rho\alpha}S^\lambda_{~\alpha}S^\beta_{~\sigma}\delta g'_{\beta\lambda}\nn\\
&=\Big(\delta^\lambda_{~\mu}S^\sigma_{~\nu}
+\delta^\lambda_{~\nu}S^\sigma_{~\mu}\Big)\delta g'_{\sigma\lambda}\,.
\end{align}
Here, $\frac{\delta \gmn}{\delta (S^{-1})^\rho_{~\sigma}}|_f$ denotes the variation of $\gmn=f_{\mu\alpha}[S^{-2}]^\alpha_{~\nu}$ with respect to $[S^{-1}]^\rho_{~\sigma}$ taken at constant $\fmn$.
The new variables are designed such that we obtain a simple expression for the variation of the inverse square-root matrix,
\beqn
\left.\frac{\delta (S^{-1})^\rho_{~\sigma}}{\delta \gmn}\right|_f\delta\gmn=f^{\rho\alpha}S^\lambda_{~\alpha}S^\beta_{~\sigma}\delta g'_{\beta\lambda}\,,
\eeqn
which can be used to obtain the variation of the square-root (from $\delta S=-S\delta S^{-1} S$)
\beqn\label{relgg}
\left.\frac{\delta S^\rho_{~\sigma}}{\delta \gmn}\right|_f\delta\gmn&=&-(S^2)^\beta_{~\sigma}g^{\rho\lambda}\delta g'_{\beta\lambda}\,.
\eeqn
In this last form we can shed some more light on this field redefinition. Consider the general variation of the squared matrix $\delta S^2$, using matrix notation,
\be\label{dS2}
\delta S^2 = \delta_gS^2+\delta_f S^2 = -g^{-1}\delta g S^2 + g^{-1}\delta f\,,
\ee
where $\delta_gS^2$ and $\delta_fS^2$ denote the variations with $\fmn$ and $\gmn$ held fixed respectively (i.e.~the subscript labels the field which is to be varied and in terms of the above notation we have that $\delta_gS^2=\left.\tfrac{\delta S^2}{\delta g}\right|_f\delta g$) and we have used the defining relation $S^2=g^{-1}f$. Focussing on the $\delta_gS^2$ part, the redefined fluctuations are taylored such that the expression for $\delta_gS$ in terms of $\delta g'$ has exactly the same form as $\delta_gS^2$ in terms of $\delta g$ (to wit:~$\delta_gS^2=-g^{-1}\delta gS^2$), i.e.~we can equivalently define the redefinition through (cf.~\eqref{relgg})
\be\label{dgp_def2}
\delta_g S\equiv -g^{-1}\delta g' S^2\,.
\ee
Using $\delta_g S^2=S\delta_g S+\delta_g S S$ we then obtain
\be
g^{-1}\delta g = S g^{-1}\delta g'+g^{-1}\delta g' S\,.
\ee
This expression has two immediate virtues. Firstly, by considering the component form we straightforwardly arrive at the right hand side of the definition \eqref{dgp_def1}  (using that $(gS)^{\mathrm{T}}=gS$). Secondly, we recognise this as a Sylvester equation of exactly the same form as in \eqref{dS2sylvester} and therefore deduce that $\delta \gmn'$ can be uniquely determined in terms of $\delta\gmn$ if and only if the spectra of eigenvalues for $S$ and $-S$ do not intersect, i.e.~exactly when the variation of the square-root exists and has a unique solution in terms of $\delta \gmn$ and $\delta \fmn$. This means that under these quite general conditions we are assured that the field redefinition is invertible and well defined and whenever this is not the case even the original problem of linearisation is ill-defined.

Motivated by the interchange symmetry \eqref{intsym}, we redefine the $\delta\fmn$ fluctuations mirroring the previous discussion. This leads us to the definition
\be\label{dfp_def1}
\delta_f S^{-1}\equiv-f^{-1}\delta f'S^{-2}\,,
\ee 
from which it follows that
\be
f^{-1}\delta f=S^{-1}f^{-1}\delta f'+f^{-1}\delta f'S^{-1}\,.
\ee
This is again a Sylvester equation which implies that a unique solution exist and the transformation is uniquely invertible if and only if the spectra of eigenvalues of $S^{-1}$ and $-S^{-1}$ do not intersect, which is of course equivalent to the statement that $\sigma(S)\cap\sigma(-S)=\emptyset$. In components this redefinition amounts to
\be\label{dfp_def2}
\delta\fmn=\Big(\delta^\lambda_{~\mu}[S^{-1}]^\sigma_{~\nu}
+\delta^\lambda_{~\nu}[S^{-1}]^\sigma_{~\mu}\Big)\delta f'_{\sigma\lambda}\,.
\ee
Finally we can combine these fluctuations as $\delta S=\delta_gS+\delta_fS$, 
\be\label{dS_general}
\delta S = -g^{-1}\delta g'S^2+S^{-1}g^{-1}\delta f'S^{-1}\,,
\ee
with a corresponding expression for the inverse $\delta S^{-1}=-S^{-1}\delta SS^{-1}$
\be\label{dSI_general}
\delta S^{-1} = -f^{-1}\delta f'S^{-2}+Sf^{-1}\delta g'S\,,
\ee
which, due to our definitions, can also be obtained from $\delta S$ by interchanging $\gmn$ and $\fmn$. This kind of symmetric redefinition is useful because it allows us to study one set of the equations of motion separately, e.g.~the $\gmn$ equations, and immediately deduce the corresponding results in the $\fmn$ equations due to the interchange symmetry present in the theory. We stress that i) these redefinitions are tailored such that they are interchange symmetric but also exist and are uniquely invertible under the exact same conditions for which the linearised problem is well-defined to start with and ii) the variation of the square-root is reduced to at most two terms (i.e.~one term for massive gravity and two terms when both metrics are dynamical) when using these redefined fluctuations as compared to the more than 30 terms (or even twice that if we consider bimetric theory) when using the original fluctuations. This last point makes many problems analytically tractable and is therefore potentially very important as we discuss further in section \ref{sec: conclusions}.

\section{Linear equations around general backgrounds}\label{sec: linearisation}
The virtue of the field redefinitions \eqref{dgp_def1} and \eqref{dfp_def2} is easy to understand since they circumvent the necessity of dealing with the large number of terms contained in the general solution for $\delta S$ given in \eqref{sylvestersol}. In fact, we can directly insert the expression \eqref{dS_general} in the expansions \eqref{dVmn} and \eqref{dtVmn} without any further work and simply expand the sums in these terms. 

Of course, a slight drawback is that, by performing the above redefinitions, we introduce extra contributions to the kinetic terms since the kinetic operators now act on the background fields appearing in the redefinitions. Although this might appear as an obstacle at first sight, for most applications the background is known and it usually presents no problem to compute its derivatives. Moreover, such derivatives will play no role in the subsequent analysis of constraints since the latter only depend on the terms generated by derivatives acting on the linear perturbations. In those cases, the extra terms will not cause any difficulties and these terms are far less in number and can presumably be dealt with much more easily than the variation of the square-root in the original variables. For analysing the constraint structure the redefinitions certainly prove very useful, as we will see in section \ref{sec: covconstr}.

In the remainder of this section we will simply collect all the relevant linearised expressions in terms of the redefined fluctuations.

\subsection{The linearised Einstein tensors}
Recalling the expressions \eqref{dGmn}-\eqref{fEinstOp} for the linearised Einstein tensors we trivially find that in terms of the redefined fluctuations \eqref{dgp_def1} and \eqref{dfp_def2} these now read
\begin{align}\label{kinlingp}
\delta\mathcal{G}_{\mu\nu}
=&-\tfrac{1}{2}\Big[\delta^\rho_{\mu}\delta^\sigma_{\nu}\nabla^2+g^{\rho\sigma}\nabla_\mu\nabla_\nu
-\delta^\rho_{\mu}\nabla^\sigma\nabla_\nu\nn\\
&-\delta^\rho_{\nu}\nabla^\sigma\nabla_\mu
-\gmn g^{\rho\sigma}\nabla^2
+\gmn\nabla^\rho\nabla^\sigma\nn\\
&+\delta^\rho_{\mu}\delta^\sigma_{\nu}R-\gmn R^{\rho\sigma}\Big]
\Big[\Big(\delta^\alpha_{\rho}S^\beta_{~\sigma}+\delta^\alpha_{\sigma}S^\beta_{~\rho}\Big)\delta g'_{\alpha\beta}\Big]\,,
\end{align}
where the operator in the left bracket acts on all terms in the right bracket. Moreover, from the nonlinear equations of motion~\eqref{eombim} we can derive the following background relation, expressing the above combination of curvatures in terms of $S$ (with an analogous expression for $\tilde{R}_{\mu\nu}$ in terms of $\tilde{V}_{\mu\nu}$),
\begin{align}\label{Ecomb}
\delta^\rho_\mu\delta^\sigma_\nu R-\gmn R^{\rho\sigma}=
m^2\bigl(&\delta^\rho_\mu\delta^\sigma_\nu g^{\lambda\omega}V_{\lambda\omega}
+\gmn V^{\rho\sigma}\nn\\
&-\tfrac1{2}\gmn g^{\rho\sigma}g^{\lambda\omega}V_{\lambda\omega}\bigr)\,.
\end{align}
Using this to replace the curvature terms in (\ref{kinlingp}) by interaction contributions we find that,
\begin{align}\label{lingET}
\delta\mathcal{G}_{\mu\nu}=&\,
{\mathcal{E}}_{\mu\nu}^{\ph\mu\ph\nu\rho\sigma}\left[\Big(\delta^\alpha_{\rho}S^\beta_{~\sigma}
+\delta^\alpha_{\sigma}S^\beta_{~\rho}\Big)\delta g'_{\alpha\beta}\right]\nn\\
&+\tfrac{m^2}{4}\biggl[\gmn g^{\rho\sigma}g^{\lambda\omega}V_{\lambda\omega}-2\gmn V^{\rho\sigma}\nn\\
&\quad
-2\delta^\rho_\mu\delta^\sigma_\nu g^{\lambda\omega}V_{\lambda\omega}\biggr]
\Big(\delta^\alpha_{\rho}S^\beta_{~\sigma}
+\delta^\alpha_{\sigma}S^\beta_{~\rho}\Big)\delta g'_{\alpha\beta}\,.
\end{align}
Similar manipulations lead to a corresponding expression for the variation of the $\fmn$ Einstein tensor,
\begin{align}\label{linfET}
\delta\tilde{\mathcal{G}}_{\mu\nu}=&\,
\tilde{\mathcal{E}}_{\mu\nu}^{\ph\mu\ph\nu\rho\sigma}\left[\Big(\delta^\alpha_{\rho}[S^{-1}]^\beta_{~\sigma}
+\delta^\alpha_{\sigma}[S^{-1}]^\beta_{~\rho}\Big)\delta f'_{\alpha\beta}\right]\nn\\
&+\tfrac{m^2}{4\alpha^2}\biggl[\fmn f^{\rho\sigma}f^{\lambda\omega}\tilde{V}_{\lambda\omega}
-2\fmn \tilde{V}^{\rho\sigma}\nn\\
&\quad
-2\delta^\rho_\mu\delta^\sigma_\nu f^{\lambda\omega}\tilde{V}_{\lambda\omega}\biggr]
\Big(\delta^\alpha_{\rho}[S^{-1}]^\beta_{~\sigma}
+\delta^\alpha_{\sigma}[S^{-1}]^\beta_{~\rho}\Big)\delta f'_{\alpha\beta}\,.
\end{align}
As we already mentioned in the beginning of this section, the only possible drawback of these field redefinitions is the fact that the kinetic operators now act on both the fluctuations and the background fields. 

Before ending this exposition of the linearised Einstein tensors we provide the linearised identities implied by the conservation of the Bianchi identities, i.e.,
\be
\delta\left(g^{\mu\rho}\nabla_\rho\mathcal{G}_{\mu\nu}\right)=0\,,\qquad
\delta\left(f^{\mu\rho}\tilde{\nabla}_\rho\tilde{\mathcal{G}}_{\mu\nu}\right)=0\,.
\ee
These imply the following linear identities
\begin{align}\label{linBianchi_g1}
g^{\mu\rho}\nabla_\rho\delta\mathcal{G}_{\mu\nu}=&\,
\delta g_{\mu\rho}\nabla^\rho\mathcal{G}^\mu_{~\nu}
+\mathcal{G}^\sigma_{~\nu}\nabla^\rho\delta g_{\sigma\rho}\nn\\
&-\tfrac{1}{2}g^{\lambda\rho}\mathcal{G}_{\nu\sigma}\nabla^\sigma\delta g_{\lambda\rho}
+\tfrac{1}{2}\mathcal{G}^{\sigma\mu}\nabla_\nu\delta g_{\mu\sigma}\,,
\end{align}
and similarly
\begin{align}\label{linBianchi_f1}
f^{\mu\rho}\tilde{\nabla}_\rho\delta\tilde{\mathcal{G}}_{\mu\nu}=&\,
\delta f_{\mu\rho}\tilde{\nabla}^\rho\tilde{\mathcal{G}}^\mu_{~\nu}
+\tilde{\mathcal{G}}^\sigma_{~\nu}\tilde{\nabla}^\rho\delta f_{\sigma\rho}\nn\\
&-\tfrac{1}{2}f^{\lambda\rho}\tilde{\mathcal{G}}_{\nu\sigma}\tilde{\nabla}^\sigma\delta f_{\lambda\rho}
+\tfrac{1}{2}\tilde{\mathcal{G}}^{\sigma\mu}\tilde{\nabla}_\nu\delta f_{\mu\sigma}\,.
\end{align}
Switching to the new fluctuation variables and using the background equations \eqref{eombim} to replace the Einstein tensors by interaction contributions on the right-hand side, we can write these identities as
\begin{align}\label{linBianchi_g2}
\nabla^\mu\delta\mathcal{G}_{\mu\nu}=&\,
\tfrac{m^2}{2}\Big[g^{\mu\rho}V_{\nu\sigma}\nabla^\sigma-V^{\mu\rho}\nabla_\nu
-2V^\mu_{~\nu}\nabla^\rho\nn\\
&-2(\nabla^\mu V^\rho_{~\nu})\Big]\left[
\left(\delta^\alpha_{\rho}S^\beta_{~\mu}
+\delta^\alpha_{\mu}S^\beta_{~\rho}\right)\delta g'_{\alpha\beta}\right]\,,
\end{align}
\begin{align}\label{linBianchi_f2}
\tilde\nabla^\mu\delta\tilde{\mathcal{G}}_{\mu\nu}=&\,
\tfrac{m^2}{2\alpha^2}\Big[f^{\mu\rho}\tilde{V}_{\nu\sigma}\tilde{\nabla}^\sigma
-\tilde{V}^{\mu\rho}\tilde{\nabla}_\nu
-2\tilde{V}^\mu_{~\nu}\tilde{\nabla}^\rho\nn\\
&-2(\tilde{\nabla}^\mu \tilde{V}^\rho_{~\nu})\Big]\left[
\left(\delta^\alpha_{\rho}[S^{-1}]^\beta_{~\mu}
+\delta^\alpha_{\mu}[S^{-1}]^\beta_{~\rho}\right)\delta f'_{\alpha\beta}\right]\,,
\end{align}
where again operators in the leftmost brackets act on all objects within the rightmost brackets. We note that $\nabla^\mu$ acts also on $V^\rho_{~\nu}$ in the last terms of the leftmost brackets. These expressions will be required later on when we analyse the constraint in section \ref{sec: covconstr}. We remind the reader that in these expressions indices are raised using $\gmn$ in e.g.~\eqref{lingET}, \eqref{linBianchi_g1} and \eqref{linBianchi_g2} while they are raised using $\fmn$ in \eqref{linfET}, \eqref{linBianchi_f1} and \eqref{linBianchi_f2}. This reminder is purely for the readers convenience and is consistent with our remarks in the introduction.

\subsection{The linearised interaction contributions}
The linearised interaction contributions are obtained by simply inserting \eqref{dgp_def1}, \eqref{dfp_def2} and \eqref{dS_general} into \eqref{dVmn} and \eqref{dtVmn}. This results in
\begin{widetext}
\begin{align}\label{dVmnp}
\delta V_{\mu\nu} = &\,g^{\rho\sigma}V_{\sigma\nu}\left(\delta^\alpha_{\rho}S^\beta_{~\mu}
+\delta^\alpha_{\mu}S^\beta_{~\rho}\right)\delta g'_{\alpha\beta}\nn\\
&\,-g_{\mu\rho}\sum_{n=1}^{3}(-1)^{n}\beta_n
\sum_{k=1}^{n}(-1)^{k}\biggl\{\left[S^{n-k}\right]^{\rho}_{~\nu}
\sum_{m=1}^{k}(-1)^{m}e_{k-m}(S)\left[S^{m-1}\delta S\right]^{\sigma}_{~\sigma}\nn\\
&\qquad\qquad\qquad\qquad\qquad\qquad
+e_{k-1}(S)\sum_{m=0}^{n-k}\left[S^{m}\delta S S^{n-k-m}\right]^{\rho}_{~\nu}
\biggr\}\,,
\end{align}
and
\begin{align}\label{dtVmp}
\delta \tilde{V}_{\mu\nu} =&\, f^{\rho\sigma}\tilde{V}_{\sigma\nu}\left(\delta^\alpha_{\rho}[S^{-1}]^\beta_{~\mu}
+\delta^\alpha_{\mu}[S^{-1}]^\beta_{~\rho}\right)\delta f'_{\alpha\beta}\nn\\
&\,-f_{\mu\rho}\sum_{n=1}^{3}(-1)^{n}\beta_{4-n}
\sum_{k=1}^{n}(-1)^{k}\biggl\{\left[S^{k-n}\right]^{\rho}_{~\nu}
\sum_{m=1}^{k}(-1)^{m}e_{k-m}(S^{-1})\left[S^{-m+1}\delta S^{-1}\right]^{\sigma}_{~\sigma}\nn\\
&\qquad\qquad\qquad\qquad\qquad\qquad
+e_{k-1}(S^{-1})\sum_{m=0}^{n-k}\left[S^{-m}\delta S^{-1} S^{m+k-n}\right]^{\rho}_{~\nu}
\biggr\}\,,
\end{align}
\end{widetext}
where now $\delta S$ and $\delta S^{-1}$ are given by 
\begin{align}
\delta S &= -g^{-1}\delta g'S^2+S^{-1}g^{-1}\delta f'S^{-1}\,,\\
\delta S^{-1} &= -f^{-1}\delta f'S^{-2}+Sf^{-1}\delta g'S\,.
\end{align}
Combining these expressions with the kinetic terms \eqref{lingET} and \eqref{linfET} it is now straightforward to write down the full perturbative equations \eqref{lineoms}. These can then be used for any analysis concerning linear fluctuations around some background solution of the bimetric theory.

\section{Constraint analysis}\label{sec: covconstr}
In this section we discuss the nature of constraints of bimetric theory and massive gravity at the Lagrangian level, or at the level of equations of motion. In order to do this in a self-consistent manner we first review the methodology of a covariant constraint analysis in the Lagrangian formalism within the bimetric setup. This helps to summarise the steps involved and to set up some notation for later purposes. A detailed analysis along these lines was performed in a subset of massive gravity models in \cite{Bernard:2015mkk} and is summarised in \cite{Bernard:2014bfa}. We will see that the method developed in the first part of the paper in order to simplify the linearised equations will turn out to be very useful in order to extract the constraints we are after.

\subsection{Constant curvature backgrounds}\label{sec: propconstr}
First, let us remind how the constraint analysis works out for the massive Fierz-Pauli equations on constant curvature backgrounds, i.e.~Einstein spacetimes. Note that this can equally be viewed as a bimetric analysis restricted to the proportional backgrounds of section \ref{sec:Einstein}. There we obtained one equation, i.e.~\eqref{FPeqsMeq0}, describing the propagation of a massless spin-2 field and the counting of degrees of freedom there proceeds as in linearised general relativity and gives the standard 2 polarisations. We also obtained the equation \eqref{FPeqsMneq0} which exactly coincides with the massive Fierz-Pauli equation, namely
\begin{align}\label{FPeqsMneq0_2}
\delta E_{\mu\nu}=
{\mathcal{E}}_{\mu\nu}^{\ph\mu\ph\nu\rho\sigma}\delta M_{\rho\sigma}
&-\Lambda\left(\delta M_{\mu\nu}-\frac1{2}\gmn g^{\rho\sigma}\delta M_{\rho\sigma}\right)\nn\\
&+\frac{m_{\mathrm{FP}}^2}{2}\left(\delta M_{\mu\nu}
-\gmn g^{\rho\sigma}\delta M_{\rho\sigma}\right)=0\,,
\end{align}
where we recall that (with $\nabla_\mu$ associated to the background field $\gmn$)
\begin{align}\label{gEinstOp_2}
\mathcal{E}_{\mu\nu}^{\ph\mu\ph\nu\rho\sigma}\delta M_{\rho\sigma}
\equiv &\,-\tfrac1{2}\biggl[\delta^\rho_\mu\delta^\sigma_\nu\nabla^2+g^{\rho\sigma}\nabla_\mu\nabla_\nu 
-\delta^\rho_\mu\nabla^\sigma\nabla_\nu\nn\\
&-\delta^\rho_\nu\nabla^\sigma\nabla_\mu
-g_{\mu\nu} g^{\rho\sigma}\nabla^2 +g_{\mu\nu}\nabla^\rho\nabla^\sigma\biggr]\delta M_{\rho\sigma}\,.
\end{align}
Due to the Bianchi identities obeyed by the Einstein tensor (including the $\Lambda$ terms) a divergence of these field equations yields
\begin{align}\label{trMconstr}
\nabla^\mu\delta E_{\mu\nu}&=\frac{m_{\mathrm{FP}}^2}{2}\left(
\nabla^\mu\delta M_{\mu\nu}-g^{\rho\sigma}\nabla_\nu\delta M_{\rho\sigma}\right)\,,
\\
\nabla^\mu\delta M_{\mu\nu}&=g^{\rho\sigma}\nabla_\nu\delta M_{\rho\sigma}\,,
\end{align}
where the last equality is deduced from the first on-shell and constitute 4 non-dynamical constraint equations for $\delta M_{\mu\nu}$. Taking a second divergence of the equations yields
\be\label{propdiv}
\nabla^\mu\nabla^\nu\delta E_{\mu\nu}=\frac{m_{\mathrm{FP}}^2}{2}\left(
\nabla^\mu\nabla^\nu\delta M_{\mu\nu}-g^{\rho\sigma}\nabla^2\delta M_{\rho\sigma}\right)\,.
\ee
Tracing instead the field equations \eqref{FPeqsMneq0_2} we get
\begin{align}\label{proptrace}
g^{\mu\nu}\delta E_{\mu\nu}=&\,
g^{\mu\nu}\nabla^2\delta M_{\mu\nu}-\nabla^\mu\nabla^\nu\delta M_{\mu\nu}\nn\\
&+\left(\Lambda-\frac{3m_{\mathrm{FP}}^2}{2}\right)g^{\mu\nu}\delta M_{\mu\nu}\,.
\end{align}
Hence we find that the linear combination
\begin{align}\label{FPcomb}
2\nabla^\mu\nabla^\nu\delta E_{\mu\nu}&+
m_{\mathrm{FP}}^2g^{\mu\nu}\delta E_{\mu\nu}\nn\\
&\qquad=
\frac{m_{\mathrm{FP}}^2}{2}\left(2\Lambda-3m_{\mathrm{FP}}^2\right)g^{\mu\nu}\delta M_{\mu\nu}\,,
\end{align}
constitutes another on-shell scalar constraint,\footnote{As is evident from \eqref{FPcomb} a caveat to this appears when the parameters satisfy $2\Lambda=3m_{\mathrm{FP}}^2$, saturating the so called Higuchi bound \cite{Higuchi:1986py}. In this case the left hand side vanishes identically even off-shell, such that a new linear scalar gauge symmetry emerges and the theory propagates a partially massless spin-2 field with only 4 degrees of freedom \cite{Deser:1983mm,Deser:2001us}.} $g^{\mu\nu}\delta M_{\mu\nu}=0$, which then furthermore using \eqref{trMconstr} implies $\nabla^\mu\delta M_{\mu\nu}=0$. Together these non-dynamical equations correspond to 5 constraints which can be used to remove 5 degrees of freedom from the original 10 components of the symmetric tensor perturbation $\delta M_{\mu\nu}$. Implementing the constraints, the field equations can thus be reduced to the equivalent system of equations,
\begin{align}
&\left(\nabla^2-m_{\mathrm{FP}}^2\right)\delta M_{\mu\nu}
+\tfrac{2\Lambda}{3}\,\delta M_{\mu\nu}=0\,,\nn\\
&\quad\nabla^\mu\delta M_{\mu\nu}=0\,,\qquad g^{\mu\nu}\delta M_{\mu\nu}=0\,.
\end{align}
In particular, in flat spacetime this is simply the Klein-Gordon equation for a transverse and traceless tensor field. This brief analysis shows that the massive Fierz-Pauli equation propagates 5 degrees of freedom and it follows that, for the proportional backgrounds, the bimetric theory propagates $2+5=7$ degrees of freedom.

\subsection{Constraint analysis on general backgrounds}\label{sec: covconstr_outline}

In principle, we would now like to mimic the procedure outlined in the previous subsection in the case of general background solutions, i.e.~away from the proportional backgrounds. In particular, we seek to obtain a scalar constraint from the bimetric equations which generalises the tracelessness condition $g^{\mu\nu}\delta M_{\mu\nu}=0$.

We start by outlining the general procedure for obtaining such a constraint in the full bimetric theory. This follows closely the methodology of the massive gravity analysis of~\cite{Bernard:2014bfa, Bernard:2015mkk}, with some straightforward generalisations to extend that analysis to the bimetric case. We will then make use of the fact that the bimetric theory has a massive gravity limit and that the constraint should survive this limit.\footnote{There are in fact two massive gravity limits as we mentioned in section \ref{sec: MGlim}. The limit $\alpha\rightarrow\infty$ makes $\gmn$ massive while the limit $\alpha\rightarrow0$ makes $\fmn$ massive. The constraint must therefore survive both of these limits.} This allows us to reduce the full problem by first restricting to the massive gravity limit and searching for the constraint there. As we will see the existence of the constraint in this limit is enough to determine the structure of a would be constraint in the full bimetric theory. Once we have determined this structure we return to the bimetric case and check whether it is actually a constraint or not. On the other hand, we stress from the onset that the existence of such a constraint in covariant form (i.e.~without doing a $3+1$ split) is a sufficient but not necessary condition for proving that the theory propagates the correct number of degrees of freedom.

\subsubsection{General form of the constraint in bimetric theory}

Consider again the linearised bimetric equations (cf.~\eqref{lineoms})
\begin{align}\label{bgeoms_dg2}
\delta E_{\mu\nu}&=\,
\mathcal{E}_{\mu\nu}^{\ph\mu\ph\nu\rho\sigma}\delta g_{\rho\sigma}+\tfrac1{2}\left[
g_{\mu\nu} R^{\rho\sigma}-\delta^\rho_\mu \delta^\sigma_\nu R\right]\delta g_{\rho\sigma}\nn\\
&\quad+m^2\delta V_{\mu\nu}=0\,, \\
\delta\tilde{E}_{\mu\nu}&=\,
\tilde{\mathcal{E}}_{\mu\nu}^{\ph\mu\ph\nu\rho\sigma}\delta f_{\rho\sigma}+\tfrac1{2}\left[
f_{\mu\nu} \tilde{R}^{\rho\sigma}-\delta^\rho_\mu \delta^\sigma_\nu \tilde{R}\right]\delta f_{\rho\sigma}\nn\\
&\quad+\frac{m^2}{\alpha^2}\delta\tilde{V}_{\mu\nu}=0\,.
\label{bgeoms_df2}
\end{align}
In this case we start out with {\it a priori} $10+10=20$ degrees of freedom in the symmetric fluctuations $\delta \gmn$ and $\delta \fmn$. General covariance then allows us to remove $2\times4$ degrees of freedom, being gauge redundancies or first class constraints \`a la Dirac. In addition, the Bianchi identities satisfied by the Einstein tensors allow us to find 4 vector constraints by taking a divergence of either equation (there are only 4 because of the identity \eqref{covID_1} which relates the divergences of the interaction terms), by which we can remove an additional 4 degrees of freedom. This leaves us with $20-2\times4-4=8$ degrees of freedom. In order to demonstrate covariantly that the theory only propagates $2+5=7$ degrees of freedom of a massless (2) and a massive (5) spin-2 field we need to find an additional scalar constraint in analogy with the previous section. However, we emphasise again that finding a covariant constraint in the Lagrangian formalism is a sufficient but not necessary condition for the theory to propagate 7 degrees of freedom.

Following \cite{Bernard:2015mkk} we conveniently introduce the equality symbol $\sim$ by which we will mean that two expressions are equal (off-shell) modulo terms that contain strictly less than two derivatives. For example, in this notation we can write the above equations \eqref{bgeoms_dg2} and \eqref{bgeoms_df2}
\be
\delta E_{\mu\nu}\sim \mathcal{E}_{\mu\nu}^{\ph\mu\ph\nu\rho\sigma}\delta g_{\rho\sigma}\,,\qquad
\delta\tilde{E}_{\mu\nu}\sim
\tilde{\mathcal{E}}_{\mu\nu}^{\ph\mu\ph\nu\rho\sigma}\delta f_{\rho\sigma}\,,
\ee
since the interaction terms do not contain any derivatives.
We then consider the following generalised traces of the field equations, tracing with powers of the tensor $S$
\begin{align}\label{Phi_gdef}
\Phi^g_k\equiv& [S^k]^\nu_{~\rho}\,g^{\rho\mu}\delta E_{\mu\nu}\sim
[S^k]^\nu_{~\rho}\,g^{\rho\mu}\mathcal{E}_{\mu\nu}^{\ph\mu\ph\nu\rho\sigma}\delta g_{\rho\sigma}\,,\\
\Phi^f_k\equiv& [S^k]^\nu_{~\rho}\,f^{\rho\mu}\delta \tilde{E}_{\mu\nu}\sim
[S^k]^\nu_{~\rho}\,f^{\rho\mu}\tilde{\mathcal{E}}_{\mu\nu}^{\ph\mu\ph\nu\rho\sigma}
\delta f_{\rho\sigma}\,.
\label{Phi_fdef}
\end{align}
Here it is enough to consider $k=0,1,2,3$ and still cover all possible independent traces that can be formed. This can be realised by taking into account the Cayley-Hamilton theorem, which for any $4\times4$ matrix $S$ can be stated as,
\be\label{CHthrm}
S^4-e_1(S)S^3+e_2(S)S^2-e_3(S)S+e_4(S)\mathbb{1}=0\,.
\ee
This theorem allows us to express any powers of the matrix greater than 3 in terms of a linear combination of terms with all positive but at most 3 powers of the matrix. This includes any inverse powers of the matrix since the inverse (provided that it exists) can be computed as a polynomial from \eqref{CHthrm}. Note that the case $k=0$ above corresponds to the ordinary metric traces. The definition $\fmn=g_{\mu\rho}S^\rho_{~\sigma}S^\sigma_{~\nu}$ means that tracing with the metrics interchanged is also included in the above cases. Furthermore, the background relations $\mathcal{G}_{\mu\nu}+m^2V_{\mu\nu}=0$ and $\tilde{\mathcal{G}}_{\mu\nu}+\tfrac{m^2}{\alpha^2}\tilde{V}_{\mu\nu}=0$ are linear in the curvatures and can therefore be used to solve for $R_{\mu\nu}$ and $\tilde{R}_{\mu\nu}$ as polynomials in $S$. Since the metrics, the curvatures and the square-root matrix are the only covariant objects available to trace with, this shows that the above list is indeed exhaustive. 

Along similar lines we consider the generalised divergences of the field equations
\begin{align}\label{Psi_gdef}
\Psi^g_k\equiv& [S^k]^\nu_{~\rho}\nabla^\rho\nabla^\mu\delta E_{\mu\nu}\,,\\
\Psi^f_k\equiv& [S^k]^\nu_{~\rho}\tilde{\nabla}^\rho\tilde{\nabla}^\mu\delta \tilde{E}_{\mu\nu}\,.
\label{Psi_fdef}
\end{align}
For the same reasons as outlined above we only need to consider $k=0,1,2,3$ to exhaust all possible independent terms. Here we should note that in principle it is possible to consider different combinations of covariant derivatives, e.g.~a term like $\tilde{\nabla}^\nu\nabla^\mu\delta E_{\mu\nu}$. We have omitted these since the covariant derivatives are equal up to lower order terms, i.e.~$\nabla_\mu\sim\tilde\nabla_\mu$, the difference only containing derivatives of the background fields.

Having now defined all possible scalars that can be constructed from the field equations, we consider a general linear combination,
\be\label{Cdef_gf}
\mathcal{C}\equiv \sum_{k=0}^3\left(u^g_k\Phi^g_k+v^g_k\Psi^g_k+u^f_k\Phi^f_k+v^f_k\Psi^f_k\right)\,,
\ee
where the 16 coefficients $\{u^{g,f},v^{g,f}\}$ are scalar functions of the background fields. The aim then is to determine 15 of these, since one of them can be scaled away without any loss of generality, such that $\mathcal{C}\sim0$. If this is possible then on-shell the equation $\mathcal{C}=0$ will provide the sought after covariant scalar constraint since it is a nontrivial, non-dynamical equation with no double time derivatives appearing.

\subsubsection{General form of the constraint in massive gravity}\label{sec: covconstrMG}
Due to the linearity of the problem at hand and the interchange symmetry \eqref{intsym} of the theory it is useful to split the problem into separate parts and first study only one of the bimetric equations. In more detail, consider the $\gmn$ equations written in the form
\be\label{dElinMG}
\delta E_{\mu\nu}=\delta_g\mathcal{G}_{\mu\nu}+m^2\delta_gV_{\mu\nu}
+m^2\delta_fV_{\mu\nu}=0\,,
\ee
where the subscripts on the variations denote variations with respect to the corresponding field, e.g.~$\delta_gV_{\mu\nu}=\tfrac{\p V_{\mu\nu}}{\p g_{\rho\sigma}}\delta g_{\rho\sigma}$ with $\fmn$ held fixed etc. We know from previous studies that a constraint does exist for these equations in the massive gravity limit (i.e.~for $\delta\fmn\rightarrow0$ and the last term is absent) when one uses the 
vielbein formalism \cite{Deffayet:2012nr,Deser:2014hga} as well as, for non vanishing $\beta_1$, in the metric formalism \cite{Bernard:2014bfa,Bernard:2015mkk}. The generalised St\"uckelberg analysis of~\cite{Kugo:2014hja} also implied the absence of the Boulware-Deser ghost in full generality within the metric formulation.\footnote{The analysis of~\cite{Kugo:2014hja} however relied crucially on background configurations for which $S$ could be taken to be diagonal, and hence was not fully covariant. We will come back to this point later in section~\ref{sec: MGsclrconstr}.\label{Kugo}} As discussed briefly in section \ref{sec: MGlim} we know that there exist bimetric solutions with a well-behaved massive gravity limit even at the linearised level. It is thus perfectly reasonable to assume that the constraint must survive the massive gravity limit in the metric formalism and in the most general case. In order to be able to obtain a bimetric constraint with this behaviour we can then clearly focus only on the first two terms in \eqref{dElinMG}, which are linear in $\delta\gmn$ and do not contain $\delta\fmn$. In other words, we first study the $\gmn$ equations $\delta E_{\mu\nu}$ and set the $\fmn$ perturbations to zero in that analysis. This is equivalent to considering the constraint analysis in the strict massive gravity limit $\alpha\rightarrow\infty$. Interchange symmetry (c.f.~\eqref{intsym}) then allows us to deduce corresponding conclusions for the $\fmn$ equations $\delta\tilde{E}_{\mu\nu}$ with the $\delta\gmn$ fluctuations set to zero. Finally we combine and extend our results to discuss the outcome of the analysis in the full bimetric theory.

Considering only the $\gmn$ equations, with $\delta\fmn=0$ momentarily, we set out to find a scalar constraint of the form (cf.~\eqref{Cdef_gf})
\be
\mathcal{C}_g\equiv \sum_{k=0}^3\left(u^g_k\Phi^g_k+v^g_k\Psi^g_k\right)\,.
\ee
where the $\{u^g,v^g\}$ are scalar coefficients to be determined such that $\mathcal{C}_g\sim0$, i.e.~contains only terms with less than two derivatives acting on the fluctuations. We also recall the definitions \eqref{Phi_gdef} and \eqref{Psi_gdef}
\begin{align}\label{Phi_gdef2}
\Phi^g_k\equiv& [S^k]^\nu_{~\rho}\,g^{\rho\mu}\delta E_{\mu\nu}\sim
[S^k]^\nu_{~\rho}\,g^{\rho\mu}\mathcal{E}_{\mu\nu}^{\ph\mu\ph\nu\rho\sigma}\delta g_{\rho\sigma}\,,\\
\Psi^g_k\equiv& [S^k]^\nu_{~\rho}\nabla^\rho\nabla^\mu\delta E_{\mu\nu}\,,
\label{Psi_gdef2}
\end{align}
where here and for the rest of this subsection we use $\gmn$ to raise indices everywhere. It is worth pointing out that \eqref{Phi_gdef2} also holds true in the case of nonzero $\delta\fmn$ since the 2-derivative parts of the traces only come from the kinetic operator. We proceed to give the expressions of these generalised traces and divergences in full detail.

\subsubsection{The generalised traces}\label{sec: gentrace}
From the definition of $\mathcal{E}$ in \eqref{gEinstOp} it is straightforward to find that, in general\
\begin{align}
\Phi^g_k\sim-\Big[&[S^{k+1}]^{\rho\sigma}g^{\kappa\mu}+S^{\rho\sigma}[S^k]^{\kappa\mu}\nn\\
&-S^{\sigma\kappa}[S^k]^{\rho\mu}-[S^{k+1}]^{\sigma\mu}g^{\rho\kappa}\nn\\
&-[S^{k}]^\lambda_{~\lambda}S^{\rho\sigma}g^{\kappa\mu}
+[S^k]^\lambda_{~\lambda}g^{\rho\kappa}S^{\sigma\mu}
\Big]\nabla_{\kappa}\nabla_{\mu}\delta g'_{\rho\sigma}\,.
\end{align}
The relevant ones for our purposes are given explicitly by
\begin{subequations}\label{gentrace}
\begin{align}\label{Phi_0}
\Phi^g_0&\sim-2\Big[g^{\rho\kappa}S^{\sigma\mu}
-S^{\rho\sigma}g^{\kappa\mu}
\Big]\nabla_{\kappa}\nabla_{\mu}\delta g'_{\rho\sigma}
\,,\\
\Phi^g_1&\sim-\Big[[S^{2}]^{\rho\sigma}g^{\kappa\mu}+S^{\rho\sigma}S^{\kappa\mu}\nn\\
&\quad\quad-S^{\sigma\kappa}S^{\rho\mu}-[S^{2}]^{\sigma\mu}g^{\rho\kappa}\nn\\
&\quad\quad-e_1S^{\rho\sigma}g^{\kappa\mu}
+e_1g^{\rho\kappa}S^{\sigma\mu}
\Big]\nabla_{\kappa}\nabla_{\mu}\delta g'_{\rho\sigma}
\,,\\
\Phi^g_2&\sim-\Big[[S^{3}]^{\rho\sigma}g^{\kappa\mu}+S^{\rho\sigma}[S^2]^{\kappa\mu}
-S^{\sigma\kappa}[S^2]^{\rho\mu}\nn\\
&\quad\quad-[S^{3}]^{\sigma\mu}g^{\rho\kappa}
-\left(e_1^2-2e_2\right)S^{\rho\sigma}g^{\kappa\mu}\nn\\
&\quad\quad+\left(e_1^2-2e_2\right)g^{\rho\kappa}S^{\sigma\mu}
\Big]\nabla_{\kappa}\nabla_{\mu}\delta g'_{\rho\sigma}
\,,\\
\Phi^g_3&\sim-\Big[[S^{4}]^{\rho\sigma}g^{\kappa\mu}+S^{\rho\sigma}[S^3]^{\kappa\mu}
-S^{\sigma\kappa}[S^3]^{\rho\mu}\nn\\
&\quad\quad-[S^{4}]^{\sigma\mu}g^{\rho\kappa}
-\left(e_1^3-3e_1e_2+3e_3\right)S^{\rho\sigma}g^{\kappa\mu}\nn\\
&\quad\quad+\left(e_1^3-3e_1e_2+3e_3\right)g^{\rho\kappa}S^{\sigma\mu}
\Big]\nabla_{\kappa}\nabla_{\mu}\delta g'_{\rho\sigma}\,,
\label{Phi_3}
\end{align}
\end{subequations}
where, for brevity, we have suppressed the functional dependence on $S$ in all of the $e_n(S)$ appearing here. We note that all of the $\beta_3$ terms, i.e.~the~$\Phi_3^g$ terms, in general contain 4 powers of $S$ in various forms. In particular, we observe that there are two terms which directly involve the 4th power of $S$. These could in principle, from the Cayley-Hamilton theorem (cf.~\eqref{CHthrm}), be reduced to lower orders by using $S^4=-\sum_{n=1}^{4}(-1)^{n}e_{n}(S)S^{4-n}$. We will refrain from doing this for now since, as we will find out later, it is in fact not necessary.

The traces $\Phi^f_k$ from the $\fmn$ equations can be obtained directly from these expressions by the following formal replacements\footnote{These replacements are consistent with the interchange symmetry \eqref{intsym} with $\alpha=1$. The reason we do not formally need to include any powers of $\alpha$ here is that, since the $\Phi^g_k$  ($\Phi^f_k$) only contain terms of a single degree in $S$ ($S^{-1}$), any such powers of $\alpha$ simply show up as overall relative factors which can conveniently be absorbed into the definitions of the coefficients $\{u^{g,f},v^{g,f}\}$ for our purposes.}
\begin{align}
&S^{\rho}_{~\nu}\rightarrow[S^{-1}]^{\rho}_{~\nu}\,,\quad e_n(S)\rightarrow e_n(S^{-1})\,,\nn\\
&\gmn\rightarrow\fmn\,,\quad\delta\gmn'\rightarrow\delta\fmn'\,,
\end{align}
and correspondingly considering all index raisings to be done with respect to $\fmn$ (the covariant derivatives need not be replaced since when acting on the fluctuations they differ only by terms $\sim0$).

At this level the field redefinition \eqref{dgp_def1} has not simplified the expressions as compared to using the original fluctuation fields $\delta\gmn$. In fact, each individual $\Phi^g_k$ contains as many terms and even an extra power of $S$. 
Moreover, as we saw in section \ref{sec: propconstr}, around the proportional backgrounds, $\fmn=c^2\gmn$, the two-derivative terms in the trace of the linearised equations \eqref{proptrace}, which cancelled out against the two-derivative terms in the double divergence \eqref{propdiv}, were of the form $(g^{\mu\nu}\nabla^2-\nabla^\mu\nabla^\nu)\delta\gmn$. On the same backgrounds, we have that $S^\mu_{~\nu}=c\,\delta^\mu_{\nu}$ and hence all of the $\Phi^g_k$ above simply reduce to functions proportional to $(g^{\mu\nu}\nabla^2-\nabla^\mu\nabla^\nu)\delta\gmn$. From this observation we deduce that all of the $\Phi^g_k$ could potentially contribute to the constraint. 
In the next section we identify the correct combination of generalised traces through examining the double divergences $\Psi^g_k$ which involve the variation of the square-root matrix. In this context, the advantage of working with the redefined fluctuations will become evident.

\subsubsection{The generalised divergences}\label{sec: gendiv}
In order to treat the generalised divergences systematically we first define and compute the tensor
\be
\Psi^{\lambda}_{~\nu}\equiv \nabla^{\lambda}\nabla^{\mu}\delta E_{\mu\nu}\,.
\ee
The generalised divergences $\Psi^g_k$ are then simply obtained by tracing this with powers of $S$, i.e.~ $\Psi^g_k=[S^k]^\nu_{~\lambda}\Psi^{\lambda}_{~\nu}$. Using the explicit expansion of $\delta V_{\mu\nu}$ provided in eq.~\eqref{dVmnp}, together with the linearised Bianchi identity given in \eqref{linBianchi_g2} we find that $\Psi^{\lambda}_{~\nu}$ can be written on the form
\begin{widetext}
\begin{align}\label{Psimn}
\Psi^{\lambda}_{~\nu}\sim m^2\Bigg\{&
\beta_1\left[g^{\mu\rho}[S^2]^{\sigma}_{~\nu}-S^{\rho\sigma}S^\mu_{~\nu}\right]\nn\\
&+\beta_2\left[S^\mu_{~\nu}[S^2]^{\rho\sigma}+S^{\rho\sigma}[S^2]^\mu_{~\nu}
-g^{\mu\rho}[S^3]^\sigma_{~\nu}-S^{\mu\rho}[S^2]^\sigma_{~\nu}
+e_1g^{\mu\rho}[S^2]^\sigma_{~\nu}-e_1S^{\rho\sigma}S^\mu_{~\nu}\right]\nn\\
&
+\beta_3\Bigl[g^{\mu\rho}[S^4]^\sigma_{~\nu}-S^\mu_{~\nu}[S^3]^{\rho\sigma}
-S^{\rho\sigma}[S^3]^\mu_{~\nu}+S^{\mu\rho}[S^3]^\sigma_{~\nu}
+[S^2]^{\mu\rho}[S^2]^\sigma_{~\nu}-[S^2]^\mu_{~\nu}[S^2]^{\rho\sigma}\nn\\
&\quad\quad\quad +e_1S^\mu_{~\nu}[S^2]^{\rho\sigma}+e_1S^{\rho\sigma}[S^2]^\mu_{~\nu}
-e_1g^{\mu\rho}[S^3]^\sigma_{~\nu}-e_1S^{\mu\rho}[S^2]^\sigma_{~\nu}\nn\\
&\quad\quad\quad +e_2g^{\mu\rho}[S^2]^\sigma_{~\nu}-e_2S^{\rho\sigma}S^\mu_{~\nu}
\Bigr]
\Bigg\}\nabla^\lambda\nabla_\mu\delta g'_{\rho\sigma}\,,
\end{align}
\end{widetext}
where all the $e_n$ are functions of $S$. By simple inspection of this expression we make a couple of immediate and important observations:
\begin{itemize}
\item
The $\beta_1$ terms are quite simple and generically a 2nd order polynomial in $S$. Additionally, there is always a loose index on one of the $S$ appearing in these terms so that they can immediately be brought to the same form as $\Phi^g_0$ by tracing $\Psi^{\lambda}_{~\nu}$ with the inverse of $S$, i.e.~by considering $\bar{\Psi}\equiv[S^{-1}]^\nu_{~\lambda}\Psi^{\lambda}_{~\nu}$. This follows since the covariant derivatives commute up to terms $\sim0$ and means that by mere inspection we can derive the constraint for the $\beta_1$ model (with $\beta_2=\beta_3=0$) as given by $\mathcal{C}_g=\tfrac{1}{2}\beta_1\Phi_0^g+\tfrac1{m^2}\bar{\Psi}\sim0$. This constraint is exactly the same as the one identified in~\cite{Bernard:2014bfa,Bernard:2015mkk}, as can easily be seen by using the Cayley-Hamilton theorem \eqref{CHthrm} to evaluate $S^{-1}=\tfrac{1}{e_4}(e_3\mathbb{1}-e_2 S+e_1S^2-S^3)$. The simplification of the analysis performed here should however not be underestimated. To reiterate, in order to find a constraint for the $\beta_1$ model, here we only needed to match two expressions, each containing two independent terms, by simple inspection.

\item
Secondly, after considering the first point above and realising that also all of the $\beta_2$ terms have a loose index on one of the $S$ appearing in these terms, it is immediately recognised that if we now trace $\Psi^{\lambda}_{~\nu}$ with the inverse of $S$ in the general case, the $\beta_2$ terms become proportional to $\Phi^g_1$. We can thus infer, again by simple inspection, that the same contraction of $\Psi^{\lambda}_{~\nu}$, i.e.~$[S^{-1}]^\nu_{~\lambda}\Psi^{\lambda}_{~\nu}$, will enter the scalar constraint in the case when only $\beta_3=0$.

\item
Thirdly, the $\beta_3$ terms contain various contributions which are generically 4th power in $S$ but they can also all be reduced one order by tracing with $S^{-1}$. It is, however, not obvious if one can cancel the remaining terms against any of the traces $\Phi^g_k$. In fact, a computer-supported analysis shows that this is not possible and hence that a covariant constraint cannot be obtained when $\beta_3\neq 0$. In more detail, we have constructed syzygies (independent algebraic identities) along the lines of~\cite{Bernard:2015mkk} and performed a computer based analysis to check whether by implementing these it is possible to cancel the two-derivative terms of $\Phi_k^g$ against those of $\bar{\Psi}$ (or in fact any other contraction of $\Psi^\lambda_{~\nu}$). This analysis reveals that this is not possible.

\end{itemize}
Considering all of the above points it is quite obvious that the only possible scalar contraction of $\Psi^{\lambda}_{~\nu}$ we need to consider is $\bar{\Psi}\equiv[S^{-1}]^\nu_{~\lambda}\Psi^{\lambda}_{~\nu}$. The detailed expression for this contraction can be written, 
\begin{widetext}
\begin{align}\label{barPsi}
\bar{\Psi}\sim m^2\Bigg\{&
\beta_1\Bigl[g^{\rho\kappa}S^{\sigma\mu}-S^{\rho\sigma}g^{\mu\kappa}\Bigr]\nn\\
&+\beta_2\Bigl[g^{\mu\kappa}[S^2]^{\rho\sigma}+S^{\rho\sigma}S^{\mu\kappa}
-g^{\rho\kappa}[S^2]^{\sigma\mu}-S^{\mu\rho}S^{\sigma\kappa}
+e_1g^{\rho\kappa}S^{\mu\sigma}-e_1g^{\mu\kappa}S^{\rho\sigma}\Bigr]\nn\\
&
+\beta_3\Bigl[g^{\rho\kappa}[S^3]^{\mu\sigma}-g^{\mu\kappa}[S^3]^{\rho\sigma}
-S^{\rho\sigma}[S^2]^{\mu\kappa}+S^{\mu\rho}[S^2]^{\sigma\kappa}
+[S^2]^{\mu\rho}S^{\sigma\kappa}-S^{\mu\kappa}[S^2]^{\rho\sigma}\nn\\
&\hspace{60pt} +e_1g^{\mu\kappa}[S^2]^{\rho\sigma}+e_1S^{\rho\sigma}S^{\mu\kappa}
-e_1g^{\mu\rho}[S^2]^{\sigma\kappa}-e_1S^{\mu\rho}S^{\sigma\kappa}
\nn\\
&\hspace{60pt} +e_2g^{\mu\rho}S^{\sigma\kappa}-e_2S^{\rho\sigma}g^{\mu\kappa}
\Bigr]
\Bigg\}\nabla_\kappa\nabla_\mu\delta g'_{\rho\sigma}\,,
\end{align}
\end{widetext}
where again all the $e_n$ are functions of $S$. We note that, using the Cayley-Hamilton theorem \eqref{CHthrm} to re-express $S^{-1}$, this can equivalently be expressed as a linear combination of the contractions $\Psi^g_k$ given by
\be\label{barPsiCH}
\bar{\Psi}=\frac{1}{e_4}\left(e_3\Psi^g_0-e_2\Psi^g_1+e_1\Psi^g_2-\Psi^g_3\right)\,.
\ee
We now simply have to assemble the observations made above and identify the linear combination of the equations \eqref{gentrace} and \eqref{barPsi} which becomes the scalar constraint.

\subsubsection{The scalar constraint in massive gravity}\label{sec: MGsclrconstr}
Guided by the observations made in the previous subsection, we divide the discussion on the nature of the constraint into two separate cases, depending on whether the parameter $\beta_3$ is vanishing or not.

\paragraph{Models with $\beta_3=0$:}
As stated above, for the models with $\beta_3=0$ it is a simple matter of inspection of the expressions in \eqref{gentrace} and \eqref{barPsi} to find that the particular combination
\be\label{MGb30constr}
\mathcal{C}_g=\tfrac{1}{2}\beta_1\Phi^g_0+\beta_2\Phi^g_1+\tfrac{1}{m^2}\bar\Psi\sim0\,.
\ee
Consequently, the on-shell condition $\mathcal{C}_g=0$ constitutes a scalar constraint since no terms with two covariant derivatives appear in this combination of equations of motion. For the case $\beta_2=0$, this combination exactly coincides with the expression obtained in \cite{Bernard:2014bfa,Bernard:2015mkk}, which can be seen by using the form \eqref{barPsiCH} for $\bar{\Psi}$ (note that the $\Psi_k$ of those works differ by a factor of 1/2 from the definition used here). We have also verified that for the proportional backgrounds, $\fmn=c^2\gmn$, this reduces to the form \eqref{FPcomb} and therefore truly is the generalisation of that constraint which implies that perturbations on constant curvature backgrounds are traceless. This check verifies that the constraint we have found is not simply a trivial combination of the vector constraints and equations of motion.

\paragraph{Models with $\beta_3\neq0$:}
For the models with nonzero $\beta_3$ it is not as simple as mere inspection but the problem is still manageable analytically and we find that the best we can accomplish is to reduce the expression to the form
\begin{align}\label{MGb3constr}
\mathcal{C}_g&=
\frac{1}{2}\beta_1\Phi^g_0+\beta_2\Phi^g_1-\beta_3\left(\Phi^g_2-e_1\Phi^g_1+\frac1{2}e_2\Phi^g_0\right)+\frac{1}{m^2}\bar\Psi\nn\\
&\sim \beta_3\left([S^2]^{\mu\rho}S^{\sigma\kappa}-S^{\mu\kappa}[S^2]^{\rho\sigma}
\right)\nabla_\kappa\nabla_\mu\delta g'_{\rho\sigma}\,.
\end{align}
Here, for the $\beta_3$ terms, we have removed as many two derivative terms as possible by adding various generalised traces $\Phi^g_k$ in order to maximally simplify the expression.\footnote{In appendix \ref{app: kinetic} we show in detail that the kinetic terms do not contain any two time derivatives of the "lapse" and "shifts" of $\delta\gmn$. Hence any traces of the kinetic terms will also not contain any such time derivatives. Adding the traces $\Phi^g_k$ therefore will not alter any conclusions but only serve to simplify the final expressions.} At first sight then, the on-shell condition $\mathcal{C}_g=0$ does not seem to constitute a constraint in this case, at least not a covariant one, since we are left with a term proportional to $\beta_3$ which contains two covariant derivatives acting on the fluctuations. This result seems at first to contradict the conclusions of~\cite{Kugo:2014hja} which, by using a generalised St\"uckelberg analysis, claimed to see the absence of the Boulware-Deser ghost in a covariant way for any values of the $\beta_n$ parameters. As pointed out already in our footnote~\ref{Kugo}, the analysis of~\cite{Kugo:2014hja} relied on background configurations for which $S$ could be taken to be diagonal, so their conclusions are strictly only valid for such configurations. On the other hand, the fact that~\eqref{MGb3constr} does not vanish in a covariant way is fully consistent with the results of \cite{Deffayet:2012nr,Deser:2014hga}, who also found that there is no covariant constraint when $\beta_3\neq0$ in the vielbein formulation of massive gravity. In \cite{Deser:2014hga} it was however argued that the combination they constructed still constituted a constraint, and we would like to see what the corresponding statement is in our metric analysis.

In order to examine more closely the remaining two derivative expression in \eqref{MGb3constr},
\be\label{2DremainMG}
\left([S^2]^{\mu\rho}S^{\sigma\kappa}-S^{\mu\kappa}[S^2]^{\rho\sigma}
\right)\nabla_\kappa\nabla_\mu\delta g'_{\rho\sigma}\,,
\ee
it is convenient to go back to the original fluctuations $\delta\gmn$ and do a $3+1$ split \`a la ADM \cite{Arnowitt:1962hi} of these fluctuations. The reason for this is simply that, in order to separate the dynamical  from the non-dynamical terms, a time direction has to be chosen. In the original equations of motion the Einstein operator in the kinetic terms acts very simply on the lapse and shifts of $\delta\gmn$ keeping these non-dynamical, revealing that only the spatial components $\delta g_{ij}$ have dynamical equations. We provide a more detailed discussion of this standard result in appendix \ref{app: kinetic}. On the other hand, the lapse and shifts of $\delta\gmn'$ have essentially been rotated by $S$ through the field redefinition \eqref{dgp_def1}. In order to directly connect to standard results and avoid a discussion on possible configurations of $S$ which may generically mix up all of the components and obscure the standard $3+1$ treatment of the fluctuations it is therefore convenient to convert back to the original fluctuations and do the ADM split there. This conversion can be done quite straightforwardly utilising the power of the $\sim$ symbol, since the covariant derivatives commute under this symbol.\footnote{Since the partial derivatives anyway commute this is no loss in generality but merely simplifies the analysis.} It is simply a matter of commuting one power of $S$ through the covariant derivatives in \eqref{2DremainMG} and then symmetrising to see that
\begin{align}
&\left([S^2]^{\mu\rho}S^{\sigma\kappa}-S^{\mu\kappa}[S^2]^{\rho\sigma}
\right)\nabla_\kappa\nabla_\mu\delta g'_{\rho\sigma}\nn\\&\sim
\left(S^{\mu\lambda}S^{\sigma\kappa}-S^{\mu\kappa}S^{\lambda\sigma}\right)
\nabla_\kappa\nabla_\mu\left(S^{\rho}_{~\lambda}\delta g'_{\rho\sigma}\right)\nn\\
&\sim\tfrac{1}{2}\left(S^{\mu\rho}S^{\sigma\kappa}-S^{\mu\kappa}S^{\rho\sigma}\right)
\nabla_\kappa\nabla_\mu\delta g_{\rho\sigma}\,.
\end{align}
Now we can safely do a $3+1$ split, choosing the $0$ components to represent the time direction. Keeping only the terms with two time derivatives the above can then be written,
\be
\mathcal{C}_g\approx\tfrac{\beta_3}{2}\left(S^{0i}S^{0j}-S^{00}S^{ij}\right)\p_0^2\delta g_{ij}\,,
\ee
where latin indices denote spatial components and the symbol $\approx$ means equality up to terms with strictly less than two times derivatives (including e.g. second derivatives, one with respect to time and the other with respect to space direction). Thus  neither $\delta g_{00}$ (that we will call here "lapse" with some abuse of terminology) nor 
the components $\delta g_{0i}$ (that we will call here "shift") of $\delta\gmn$ appear with two time derivatives in this expression.

Let us explain why this observation is in fact sufficient to demonstrate the existence of a constraint. As follows from the (3+1) decompositions of the kinetic terms given in appendix~\ref{app: kinetic}, the equations of motion $\delta E_{\mu\nu}=0$ do not contain any double time derivatives on $\delta g_{00}$ and $\delta g_{0i}$. Hence, they completely determine the accelerations $\p_0^2\delta g_{ij}$ and can be used to express the latter in terms of quantities with less than two time derivatives.  
Now the above equation, $\mathcal{C}_g=0$, has also been shown to contain no second time derivatives of $\delta g_{00}$ or $\delta g_{0i}$. It can therefore be combined with the equations of motion to eliminate all occurrences of one of the $\p_0^2\delta g_{ij}$. The remaining dynamical equations contain only double time derivatives on five out of the six spatial components $\delta g_{ij}$. 
In this sense the equation $\mathcal{C}_g=0$ provides a constraint on the dynamical fields.

Note that the fact that bimetric theory possesses interactions with a structure which makes it possible to find a scalar combination of this nature is highly non-trivial. In other words, generic non-derivative interaction terms would produce two time derivatives of either lapse and shifts or some combination thereof, thereby exciting an additional degree of freedom in the dynamical fields $\delta g_{ij}$.

\subsubsection{The scalar constraint in bimetric theory}\label{sec: BGconstr}
In the previous section we managed to find a scalar constraint in the massive gravity limit of the theory, where we took the fluctuations $\delta\fmn\rightarrow0$ and focussed exclusively on the $\gmn$ equations. We can rephrase the results of the previous section in the following way. By fixing the values
\begin{align}\label{coeffs_dg}
u^g_0&=\frac1{2}(\beta_1-\beta_3e_2(S))\,,\quad u^g_1=\beta_2+\beta_3e_1(S)\,,\nn\\
u^g_2&=-\beta_3\,,\qquad\quad u^g_3=0\,,\nn\\
v^g_0&=\frac{e_3(S)}{m^2e_4(S)}\,,\quad v^g_1=-\frac{e_2(S)}{m^2e_4(S)}\,,\nn\\
v^g_2&=\frac{e_1(S)}{m^2e_4(S)}\,,\quad v^g_3=-\frac{1}{m^2e_4(S)}\,,
\end{align}
we found that the linear combination
\be
\mathcal{C}_g=\sum_{k=0}^3\left(u^g_k\Phi^g_k+v^g_k\Psi^g_k\right)\,,
\ee
did not contain any terms with two time derivatives acting on the lapse and shifts of $\delta\gmn$. Making use of the interchange symmetry \eqref{intsym} we can mirror those arguments for $\delta\gmn\rightarrow0$ to deduce that for the values (this has been explicitly verified as well)
\begin{align}\label{coeffs_df}
u^f_0&=\frac1{2}(\beta_3-\beta_1e_2(S^{-1}))\,,\quad u^f_1=\beta_2+\beta_1e_1(S^{-1})\,,\nn\\
u^f_2&=-\beta_1\,,\qquad\qquad u^f_3=0\,,\nn\\
v^f_0&=\frac{\alpha^2e_3(S^{-1})}{m^2e_4(S^{-1})}\,,\quad v^f_1=-\frac{\alpha^2e_2(S^{-1})}{m^2e_4(S^{-1})}\,,\nn\\
v^f_2&=\frac{\alpha^2e_1(S^{-1})}{m^2e_4(S^{-1})}\,,\quad v^f_3=-\frac{\alpha^2}{m^2e_4(S^{-1})}\,,
\end{align}
the linear combination
\be
\mathcal{C}_f=\sum_{k=0}^3\left(u^f_k\Phi^f_k+v^f_k\Psi^f_k\right)\,,
\ee
does not contain any terms with two time derivatives acting on the lapse and shifts of $\delta\fmn$. This means that the combination
\be\label{Cbimetric}
\mathcal{C}=\mathcal{C}_g+\mathcal{C}_f
=\sum_{k=0}^3\left(u^g_k\Phi^g_k+v^g_k\Psi^g_k+u^f_k\Phi^f_k+v^f_k\Psi^f_k\right)\,,
\ee
is completely determined in terms of the above values for the coefficients $\{u^{g,f},v^{g,f}\}$. We then only need to check wether imposing $\mathcal{C}=0$ actually constitutes a constraint or not in the general case, i.e.~when none of the fluctuations $\delta\gmn$ or $\delta\fmn$ are vanishing. The fact that the massive gravity limit did not always allow for the constraint to be manifest in a covariant language makes it {\it a priori} unlikely that it will be manifestly covariant in the bimetric case and already hints to the fact that the analysis may be less than straightforward.

Since we already know that most of the terms that appear in \eqref{Cbimetric} are not problematic (because the massive gravity analysis already revealed that these terms do not contain second time derivatives), the only left over terms to be studied are
\be\label{BGconstrRem}
[S^{-1}]^\nu_{~\kappa}g^{\kappa\lambda}g^{\mu\sigma}\nabla_\lambda\nabla_\sigma\delta_fV_{\mu\nu}
+S^\nu_{~\kappa}f^{\kappa\lambda}f^{\mu\sigma}\tilde{\nabla}_\lambda\tilde{\nabla}_\sigma\delta_g\tilde{V}_{\mu\nu}\,.
\ee
The first term here is simply the part of the interaction terms $\delta V_{\mu\nu}$ which contains the fluctuations $\delta\fmn$ (i.e.~$\delta_fV_{\mu\nu}$), which we ignored in the massive gravity analysis of the previous section. The second term is correspondingly the part of $\tilde{V}_{\mu\nu}$ which contains the fluctuations $\delta\gmn$ (i.e.~$\delta_g\tilde V_{\mu\nu}$). Both terms have been contracted with the appropriate derivatives and powers of $S$ which were deduced from the massive gravity analysis. These two contributions are completely independent, one containing only $\delta\gmn$ and the other one containing only $\delta\fmn$, and hence any cancellation in between these terms is generically impossible. This implies that they must both be absent of two time derivatives separately. We will now demonstrate that this is indeed the case. 

The remaining two-derivative contributions to the constraint coming from the $\delta\fmn$ variations, i.e.~the first term of \eqref{BGconstrRem}, namely $\bar{\Psi}_{f}\equiv[S^{-1}]^\nu_{~\kappa}\nabla^{\kappa}\nabla^\mu\delta_fV_{\mu\nu}$, are given by
\begin{align}
\bar{\Psi}_{f}\sim-\beta_1S^\mu_{~\rho}\Biggl[&f^{\rho\alpha}f^{\beta\kappa}-f^{\rho\kappa}f^{\alpha\beta}
\Biggr]\nabla_\kappa\nabla_\mu\delta f'_{\alpha\beta}\nn\\
-\beta_2S^\mu_{~\rho}\Biggl[&S^\rho_{~\lambda}f^{\lambda\kappa}f^{\alpha\beta}
+S^\alpha_{~\lambda}f^{\lambda\beta}f^{\rho\kappa}\nn\\
&-S^\rho_{~\lambda}f^{\lambda\alpha}f^{\beta\kappa}
-S^\beta_{~\lambda}f^{\lambda\kappa}f^{\rho\alpha}\nn\\
&+e_1\left(f^{\rho\alpha}f^{\beta\kappa}-f^{\rho\kappa}f^{\alpha\beta}\right)
\Biggr]\nabla_\kappa\nabla_\mu\delta f'_{\alpha\beta}\nn\\
-\beta_3S^\mu_{~\rho}\Biggl[&-[S^2]^\rho_{~\lambda}f^{\lambda\kappa}f^{\alpha\beta}
+[S^2]^\rho_{~\lambda}f^{\lambda\alpha}f^{\beta\kappa}\nn\\
&+[S^2]^\beta_{~\lambda}f^{\lambda\kappa}f^{\rho\alpha}
-[S^2]^\alpha_{~\lambda}f^{\lambda\beta}f^{\rho\kappa}\nn\\
&+S^\rho_{~\lambda}\left(S^\beta_{~\sigma}f^{\lambda\alpha}f^{\sigma\kappa}
-S^\alpha_{~\sigma}f^{\lambda\kappa}f^{\sigma\beta}\right)\nn\\
&+e_1\biggl(S^\rho_{~\lambda}f^{\lambda\kappa}f^{\alpha\beta}
+S^\alpha_{~\lambda}f^{\lambda\beta}f^{\rho\kappa}\nn\\
&\qquad-S^\rho_{~\lambda}f^{\lambda\alpha}f^{\beta\kappa}
-S^\beta_{~\lambda}f^{\lambda\kappa}f^{\rho\alpha}\biggr)\nn\\
&+e_2\left(f^{\rho\alpha}f^{\beta\kappa}-f^{\rho\kappa}f^{\alpha\beta}\right)
\Biggr]\nabla_\kappa\nabla_\mu\delta f'_{\alpha\beta}\,.
\end{align}
If, in analogy with the previous section, we consider the original fluctuations defined by
\begin{align}
\delta\fmn=(S^{-1})^\rho_{~\mu}\delta f'_{\rho\nu}+(S^{-1})^\rho_{~\nu}\delta f'_{\rho\mu}\,,
\end{align}
we may rewrite the above expression in the form
\be\label{extraterms}
\bar\Psi_f=\bar{\Psi}_1'+\bar{\Psi}_2'+\bar{\Psi}_3'\,,
\ee
where the different $\beta_i$-dependent pieces are given by
\begin{align}
\bar{\Psi}'_1\sim-\tfrac{1}{2}\beta_1\Biggl[&(S^{-1})^{\mu\alpha}(S^{-1})^{\beta\kappa}\nn\\
&-(S^{-1})^{\mu\kappa}(S^{-1})^{\alpha\beta}
\Biggr]\nabla_\kappa\nabla_\mu\delta f_{\alpha\beta}\,,
\end{align}
\begin{align}
\bar{\Psi}'_2\sim-\tfrac{1}{2}\beta_2\Biggl[&g^{\mu\kappa}(S^{-1})^{\alpha\beta}
+(S^{-1})^{\mu\kappa}g^{\alpha\beta}
-2(S^{-1})^{\beta\kappa} g^{\alpha\mu}
\nn\\
&+e_1(S^{-1})^{\mu\alpha}(S^{-1})^{\beta\kappa}\nn\\
&-e_1(S^{-1})^{\mu\kappa}(S^{-1})^{\alpha\beta}
\Biggr]\nabla_\kappa\nabla_\mu\delta f_{\alpha\beta}\,,
\end{align}
and
\begin{align}
\bar{\Psi}'_1\sim-\tfrac{1}{2}\beta_3\Biggl[&
-S^{\mu\kappa}(S^{-1})^{\alpha\beta}
+2S^{\mu\alpha}(S^{-1})^{\kappa\beta}\nn\\
&-S^{\alpha\beta}(S^{-1})^{\mu\kappa}
+g^{\mu\alpha}g^{\beta\kappa}
-g^{\mu\kappa}g^{\alpha\beta}\nn\\
&+e_1\biggl(g^{\mu\kappa}(S^{-1})^{\alpha\beta}
+(S^{-1})^{\mu\kappa}g^{\alpha\beta}\nn\\
&\qquad-(S^{-1})^{\beta\kappa} g^{\alpha\mu}
-(S^{-1})^{\alpha\mu} g^{\beta\kappa}\biggr)\nn\\
&+e_2(S^{-1})^{\mu\alpha}(S^{-1})^{\beta\kappa}\nn\\
&-e_2(S^{-1})^{\mu\kappa}(S^{-1})^{\alpha\beta}
\Biggr]\nabla_\kappa\nabla_\mu\delta f_{\alpha\beta}\,.
\end{align}
Here all indices are raised and contracted using $\gmn$. Again choosing the $0$ component as the time direction it is now straightforward to write down all time derivatives. For simplicity, we start by considering the double time derivatives only in the $\beta_1$ term,
\begin{align}
\bar{\Psi}_1'
&\approx -\tfrac{1}{2}\beta_1\Big[(S^{-1})^{0\alpha}(S^{-1})^{\beta0}-(S^{-1})^{00}(S^{-1})^{\alpha\beta}
\Big]\p_0^2\delta f_{\alpha\beta}\nn\\
&\approx 
 -\tfrac{1}{2}\beta_1\Big[(S^{-1})^{0i}(S^{-1})^{j0}-(S^{-1})^{00}(S^{-1})^{ij}
\Big]\p_0^2\delta f_{ij}\,.
\end{align}
This contains no double time derivatives on the lapse or shifts of $\delta\fmn$. Since this term appears with a factor $e_1(S)$ in the $\beta_2$ contribution of~(\ref{extraterms}), those terms do not contain double time derivatives on $\delta f_{00}$ and $\delta f_{i0}$ either. In the $\beta_2$ term, we therefore only need to consider, 
\begin{align}
\bar{\Psi}_2'
&\approx -\tfrac{1}{2}\beta_2\Big[g^{00}(S^{-1})^{\alpha\beta}
+(S^{-1})^{00}g^{\alpha\beta}\nn\\
&\qquad\qquad-2(S^{-1})^{\beta0} g^{\alpha0}\Big]\p_0^2\delta f_{\alpha\beta}\nn\\
&\approx
-\tfrac{1}{2}\beta_2\Big[g^{00}(S^{-1})^{ij}
+(S^{-1})^{00}g^{ij}\nn\\
&\qquad\qquad-2(S^{-1})^{i0} g^{j0}\Big]\p_0^2\delta f_{ij}\,.
\end{align}
Again this has no double time derivatives on the lapse and shifts of $\delta\fmn$. In the $\beta_3$ term, the contributions proportional to $e_1(S)$ and $e_2(S)$ are of the same form as the $\beta_1$ and $\beta_2$ terms and the only terms left to consider are,
\begin{align}
\bar{\Psi}_3'
&\approx
-\tfrac{1}{2}\beta_3\Big[-S^{00}(S^{-1})^{\alpha\beta}
+2S^{0\alpha}(S^{-1})^{0\beta}\nn\\
&\qquad\qquad-S^{\alpha\beta}(S^{-1})^{00}
+g^{0\alpha}g^{\beta0}
-g^{00}g^{\alpha\beta}\Big]\p_0^2\delta f_{\alpha\beta}\nn\\
&\approx
-\tfrac{1}{2}\beta_3\Big[-S^{00}(S^{-1})^{ij}
+2S^{0i}(S^{-1})^{0j}\nn\\
&\qquad\qquad-S^{ij}(S^{-1})^{00}
+g^{0i}g^{0j}
-g^{00}g^{ij}\Big]\p_0^2\delta f_{ij}\,.
\end{align}
This shows that $\bar{\Psi}_f=[S^{-1}]^\nu_{~\kappa}\nabla^{\kappa}\nabla^\mu\delta_fV_{\mu\nu}$ does not contain any double time derivatives acting on $\delta f_{00}$ and $\delta f_{i0}$. A completely analogous calculation (or a simple argument based on the interchange symmetry~\eqref{intsym}) shows that $\bar{\Psi}_g=S^\nu_{~\kappa}f^{\kappa\lambda}f^{\mu\sigma}\tilde{\nabla}_\lambda\tilde{\nabla}_\sigma\delta_g\tilde{V}_{\mu\nu}$ does not contain any double time derivatives on $\delta g_{00}$ or $\delta g_{i0}$. 

Now the arguments for the existence of a constraint are entirely analogous to our discussion of the non-covariant constraint in massive gravity at the end of section~\ref{sec: MGsclrconstr}.
The condition $\mathcal{C}=0$ can therefore be used to solve for one (or one combination) of the 12 dynamical fields, i.e.~$\delta g_{ij}$ or $\delta f_{ij}$. This will reduce the number of dynamical fields in the equations of motion by one to 11 and using linear diffeomorphisms we may further reduce this by four to the 7 degrees of freedom of a massless and a massive spin-2 field. We have thus established that although the equation $\mathcal{C}=0$ does not constitute a manifestly covariant constraint in the bimetric case (or even in the massive gravity limit for a non-zero $\beta_3$ parameter) it does still provide the required constraint in the sense that it is an equation which contains no double time derivatives of the lapse and shifts of the original variables and can therefore be used to eliminate one of the dynamical variables.

\section{Summary \& Discussion}\label{sec: conclusions}
We started by deriving the full perturbative equations of motion \eqref{lineoms} in the ghost-free bimetric theory, with the relevant expressions provided in \eqref{dGmn}, \eqref{dtGmn}, \eqref{dVmn} and \eqref{dtVmn}. Analysing these expressions analytically is however quite cumbersome due to the presence of a perturbative expansion of the square-root matrix $S$. Strikingly, this perturbative expansion can be written on a closed and finite polynomial form as given in \eqref{sylvestersol}, but the resulting expression contains many terms. The existence of this finite polynomial expression also crucially depends on the existence of a certain matrix inverse \eqref{Xinv}. The condition that this inverse exists is equivalent to the condition that the square-root matrix $S$ and its negative $-S$ do not have any common eigenvalues. Thus, in order for the bimetric equations to have a well-defined perturbative limit, this condition must be imposed on the theory externally. The perturbative problem is only well-defined for background configurations such that this condition is satisfied, but the full physical significance of this condition is yet to be fully understood.

In order to simplify the perturbative treatment we considered field redefinitions of the fluctuations, defined by \eqref{dgp_def2} and \eqref{dfp_def1}. These redefinitions were tailored such that i) they reduced the perturbative expression for the square-root down to a single term, ii) they were interchange symmetric between the metrics and iii) they were uniquely invertible under the exact same condition for which the perturbative expression for the square-root was well-defined. These three properties all have their virtues since they simplify many problems and guarantee the existence and invertibility of the field redefinitions. Utilising these or similar field redefinitions may potentially be of use to simplify a number of perturbative problems which arise in the bimetric theory.

In the second part of this work we studied the nature of constraints at the level of equations of motion for the first time. Considering first the massive gravity limit of the theory we found that the power of the field redefinitions discussed above allowed us to confirm and generalise previous results in an almost trivial manner. By mere inspection we were able to find a covariant scalar constraint \eqref{MGb30constr} responsible for the absence of the Boulware-Deser ghost on any background when $\beta_3=0$. This analysis therefore provides a clean and very simple alternative consistency proof for nonlinear massive gravity. In general, for non-zero $\beta_3$, we found that no covariant constraint exists in massive gravity. A component analysis in a $3+1$ split however revealed that the covariant expression \eqref{MGb3constr} still constituted a constraint on the dynamical variables, albeit explicitly non-covariant in nature. This conclusion agrees well with previous studies of massive gravity in its vielbein formulation. We then generalised these results to the full bimetric theory for the first time and found that, similarly to the massive gravity case with non-zero $\beta_3$, no manifestly covariant constraint exists. Again however, a component analysis in a $3+1$ split confirmed that the expression \eqref{Cbimetric}, with coefficients given by \eqref{coeffs_dg} and \eqref{coeffs_df}, was indeed a constraint on the dynamical variables.

It is interesting to compare our results for the massive gravity limit to those of reference~\cite{Buchbinder:1999ar}, which constructed the equations of motion for a massive spin-2 field coupled to gravity in a weak-curvature regime. Our expressions for the linearised equations in the massive gravity limit can be shown to precisely agree with the ones given in that reference (in terms of our $\delta\gmn$). To make this comparison one needs to express the background values of $S$ in terms of curvatures. This can be done via a curvature expansion along the lines of \cite{Hassan:2013pca} and truncate to first order in curvature. It should be noted however that this matching can only be done directly for $\delta\gmn$ and not for the redefined $\delta\gmn'$. The reason for this is that the redefinition involves a power of $S$ and hence becomes a curvature dependent field redefinition in this scenario. Since the analysis of~\cite{Buchbinder:1999ar} only went to linear order in curvature there is a field redefinition ambiguity at the level of fluctuations so the expressions found in~\cite{Buchbinder:1999ar} are in fact not unique.

Note that while this work was being completed, Ref.~\cite{Cusin:2015tmf} appeared on the arXiv which also studies the linearisation of bimetric theory in a quite general way. However the expressions provided in our work are obtained using a different method which is manifestly covariant. We furthermore have a careful discussion on the permissibility of linearisation, provide field redefinitions which drastically simplify all computations and we discuss the structure of constraints, such that the overlap with Ref.~\cite{Cusin:2015tmf} is not substantial.

\vspace{0.5cm}

\acknowledgments
We would like to thank Gilles Esposito-Farese and S.F.~Hassan for discussions.
The research of CD and MvS leading to these results have received funding from the ERC under the European Community's Seventh Framework Programme (FP7/2007-2013 Grant no. 307934). ASM is supported by ERC grant no.~615203 under the FP7 and the Swiss National Science Foundation through the NCCR SwissMAP.

\appendix

\section{Non-covariant constraint analysis, kinetic terms}\label{app: kinetic}
Here we provide some details concerning the standard general relativistic result that the Einstein-Hilbert kinetic operator does not contain any two time derivatives of the lapse $\delta g_{00}$ and shift components $\delta g_{0i}$ but only provide dynamical equations for the spatial components $\delta g_{ij}$ of the fluctuations $\delta\gmn$. We recall that, for any background configuration $\gmn$, the terms with two derivatives contained in the Einstein-Hilbert kinetic operator are given by
\begin{align}\label{kinterms}
\mathcal{E}_{\mu\nu}^{\ph\mu\ph\nu\rho\sigma}\delta g_{\rho\sigma}
\equiv\,  -\tfrac1{2}\biggl[&
g^{\rho\sigma}\nabla_\rho\nabla_\sigma\delta \gmn
+g^{\rho\sigma}\nabla_\mu\nabla_\nu\delta g_{\rho\sigma} \nn\\
&-g^{\rho\sigma}\nabla_\rho\nabla_\nu\delta g_{\mu\sigma}
-g^{\rho\sigma}\nabla_\rho\nabla_\mu\delta g_{\nu\sigma}\nn\\
&-g_{\mu\nu} g^{\rho\sigma}g^{\alpha\beta}\nabla_\alpha\nabla_\beta\delta g_{\rho\sigma}\nn\\
& +g_{\mu\nu}g^{\alpha\beta}g^{\rho\sigma}\nabla_\alpha\nabla_\rho\delta g_{\beta\sigma}
 \biggr]\,,
\end{align}
and similarly of course for $\delta\fmn$. We now do a $3+1$ split and choose the $0$ component to represent the temporal direction, while latin indices will represent spatial components. Dropping all terms with less than two time derivatives we then obtain
\begin{align}\label{kinterms2}
\mathcal{E}_{\mu\nu}^{\ph\mu\ph\nu\rho\sigma}\delta g_{\rho\sigma}
 \approx-\tfrac1{2}\big[&
g^{00}\p_0^2\delta \gmn
+g^{\rho\sigma}\partial_\mu\partial_\nu\delta g_{\rho\sigma}\nn\\ 
&-g^{0\sigma}\partial_0\partial_\nu\delta g_{\mu\sigma}
-g^{0\sigma}\partial_0\partial_\mu\delta g_{\nu\sigma}\nn\\
&-g_{\mu\nu} g^{\rho\sigma}g^{00}\p_0^2\delta g_{\rho\sigma}\nn\\
 &+g_{\mu\nu}g^{0\beta}g^{0\sigma}\p_0^2\delta g_{\beta\sigma}
 \big]\,.
\end{align}
It is now straightforward to see that the double time derivatives in the $00$-component of these equations are given by
\begin{align}\label{kinterms00}
\mathcal{E}_{00}^{\ph\mu\ph\nu\rho\sigma}\delta g_{\rho\sigma}
 \approx-\tfrac1{2}\big[ 
g^{ij}
-g_{00} g^{ij}g^{00}
  +g_{00}g^{0i}g^{0j}
 \big]\p_0^2\delta g_{ij}\,.
\end{align}
For the $0i$-components we have
\begin{align}\label{kinterms0i}
\mathcal{E}_{0i}^{\ph\mu\ph\nu\rho\sigma}\delta g_{\rho\sigma}
\approx-\tfrac1{2}\big[
-g^{0j}\delta^k_{~i}
-g_{0i} g^{jk}g^{00}
  +g_{0i}g^{0j}g^{0k}
 \big]\p_0^2\delta g_{jk}\,.
\end{align}
Finally, the double time derivatives contained in the $ij$-components are,
\begin{align}\label{kintermsij}
\mathcal{E}_{ij}^{\ph\mu\ph\nu\rho\sigma}\delta g_{\rho\sigma}
 \approx-\tfrac1{2}\big[
g^{00}\delta^l_{~i}\delta^k_{~j}
-g_{ij} g^{kl}g^{00}
  +g_{ij}g^{0l}g^{0k}
 \big]\p_0^2\delta g_{lk}\,.
\end{align}
We see that none of (\ref{kinterms00}), (\ref{kinterms0i}) and (\ref{kintermsij}) contain terms with double time derivatives on $\delta g_{00}$ or $\delta g_{0i}$. This is of course simply equivalent to the statement that the lapse and shifts of $\delta\gmn$ enter as non-dynamical variables when the kinetic operator is that of general relativity. For any theory where the kinetic operator is the standard general relativistic Einstein-Hilbert operator it is therefore only the spatial components $\delta g_{ij}$ which receive dynamical equations. 

Note that when doing a $3+1$ split in such a theory it is useful to consider the original variables, on which the Einstein-Hilbert operator acts, in order to simplify the problem. One may still consider field redefinitions which keep the structure of lapse and shifts intact, i.e.~dilates the lapse and rotates the shifts, without complicating the analysis. But any field redefinition which does not preserve this structure inevitably obscures an analysis in this language.


\end{document}